\documentclass{jpp}

\usepackage{graphicx}
\usepackage{natbib}

\ifCUPmtlplainloaded \else
  \checkfont{eurm10}
  \iffontfound
    \IfFileExists{upmath.sty}
      {\typeout{^^JFound AMS Euler Roman fonts on the system,
                   using the 'upmath' package.^^J}%
       \usepackage{upmath}}
      {\typeout{^^JFound AMS Euler Roman fonts on the system, but you
                   dont seem to have the}%
       \typeout{'upmath' package installed. JPP.cls can take advantage
                 of these fonts, if you use 'upmath' package.^^J}%
      }
  \else
  \fi
\fi

\ifCUPmtlplainloaded \else
  \checkfont{msam10}
  \iffontfound
    \IfFileExists{amssymb.sty}
      {\typeout{^^JFound AMS Symbol fonts on the system, using the
                'amssymb' package.^^J}%
       \usepackage{amssymb}%
       \let\le=\leqslant  
       \let\ge=\geqslant  
      }{}
  \fi
\fi

\ifCUPmtlplainloaded \else
  \IfFileExists{amsbsy.sty}
    {\typeout{^^JFound the 'amsbsy' package on the system, using it.^^J}%
     \usepackage{amsbsy}}
    {\providecommand\boldsymbol[1]{\mbox{\boldmath $##1$}}}
\fi

\newsavebox{\astrutbox}
\sbox{\astrutbox}{\rule[-5pt]{0pt}{20pt}}

\title[Dynamics of non-spherical bodies. Zeldovich's pancakes]{Regular and chaotic dynamics of non-spherical bodies. Zeldovich's pancakes and emission of very long gravitational waves}

\author[G. S. Bisnovatyi-Kogan and O. Yu. Tsupko]%
{G.\ns S.\ns B\ls I\ls S\ls N\ls O\ls V\ls A\ls T\ls Y\ls I \ls -\ls K\ls O\ls G\ls A\ls N $^{1,2}$%
  \thanks{Email address for correspondence: gkogan@iki.rssi.ru, tsupko@iki.rssi.ru}\ns
\and O.\ns YU.\ns T\ls S\ls U\ls P\ls K\ls O$^{1,2}$}

\affiliation{$^1$Space Research Institute of Russian Academy of Sciences, Profsoyuznaya 84/32, Moscow 117997, Russia\\[\affilskip]
$^2$National Research Nuclear University MEPhI (Moscow Engineering Physics Institute), Kashirskoe Shosse 31, Moscow 115409, Russia}

\begin{document}

\maketitle

\begin{abstract}
In this paper we review a recently developed approximate method for investigation
of dynamics of compressible ellipsoidal figures. Collapse and subsequent behaviour
are described by a system of ordinary differential equations for time evolution of
semi-axes of a uniformly rotating, three-axis, uniform-density ellipsoid. First, we
apply this approach to investigate dynamic stability of non-spherical bodies. We
solve the equations that describe, in a simplified way, the Newtonian dynamics
of a self-gravitating non-rotating spheroidal body. We find that, after loss of
stability, a contraction to a singularity occurs only in a pure spherical collapse,
and deviations from spherical symmetry prevent the contraction to the singularity
through a stabilizing action of nonlinear non-spherical oscillations. The development
of instability leads to the formation of a regularly or chaotically oscillating body,
in which dynamical motion prevents the formation of the singularity. We find
regions of chaotic and regular pulsations by constructing a Poincaré diagram. A
real collapse occurs after damping of the oscillations because of energy losses,
shock wave formation or viscosity. We use our approach to investigate approximately
the first stages of collapse during the large scale structure formation. The theory
of this process started from ideas of Ya. B. Zeldovich, concerning the formation
of strongly non-spherical structures during nonlinear stages of the development of
gravitational instability, known as 'Zeldovich's pancakes'. In this paper the collapse
of non-collisional dark matter and the formation of pancake structures are investigated
approximately. Violent relaxation, mass and angular momentum losses are taken into
account phenomenologically. We estimate an emission of very long gravitational
waves during the collapse, and discuss the possibility of gravitational lensing and
polarization of the cosmic microwave background by these waves.
\end{abstract}

\section{Introduction} \label{section-intro-main}

Observed Universe is non-homogeneous at different scales. Perturbations of density lead to gravitational collapse of matter and formation of structures of different sizes: stars, galaxies, large scale structure. Many of astrophysical objects are essentially non-spherical, and their formation and evolution can be approximately described as a behaviour of self-gravitating ellipsoidal bodies in the frame of Newtonian gravity.

Theory of ellipsoidal figures was developed in classical works of Newton, Maclaurin, Jacobi, Poincare, Chandrasechar. The systematic analysis of ellipsoidal figures of equilibrium is presented in the  book of \cite{cha}. Different types of incompressible ellipsoids are investigated there on a basis of
virial equations, as well as their dynamical and secular
stability.

The shape of ellipsoid is determined by sizes of its three semi-axes. Non-rotating figures always take the spherical form, all semi-axes are equal. In the presence of rotation the spheroid (axially-symmetrical ellipsoid with two equal semi-axes) is formed (Maclaurin spheroid). If the angular momentum of rotation is large enough, development of different kinds of instabilities lead to formation of a three-axis ellipsoid, with rotation around the small axis (Jacobi ellipsoid). Incompressible spheroid can be secularly or dynamically unstable for transition to three-axis ellipsoid. The points of the onset of the bar-mode dynamical, and secular instabilities of the Maclaurin spheroid, for its transition into the Jacobi ellipsoid, are found in the book \cite{cha}.

Approach based on ellipsoidal description allows to investigate the time evolution of a body considering behaviour of its semi-axes. First of all, its initial collapse can be described, which is a central problem in formation of astrophysical objects. In the case of the collapse  compressible ellipsoidal objects are considered. \citet{LB64, LB65}
applied the Chandrasechar's virial tensor method for the rotating,
self-gravitating spheroid of pressure-free gas, and had shown the
growth of non-axisymmetric perturbations during the collapse. It
was also discussed that the slowly shrinking Maclaurin spheroid
will enter the Jacobi series if it shrinks slowly enough for the
dissipative mechanisms to be operative. \citet{z1964} considered the dynamics of rotating dust ellipsoids
with linear dependence of velocities on coordinates, when the
problem is solved exactly. There are numerical
investigations of collapsing pressureless spheroids in papers of
\citet{LB64} and \citet{LMS}. In the paper of \citet{RT91} the virial equations for rotating Riemann
ellipsoids of incompressible fluid are demonstrated to form a
Hamiltonian dynamical system. There is a detailed description of
ellipsoid model of rotating stars in papers of \citet{lrs1, lrs4,
lrs5}. Using a variational principle, they had derived and
investigated equations of the evolution of a compressible
Riemann-S ellipsoid, incorporating viscous dissipation and
gravitational radiation. Solutions of these approximate equations
permitted to obtain equilibrium models, and to investigate their
dynamical and secular stability.
The point of instability of compressible Maclaurin spheroids is found in \cite{lrs1} and \cite{shap04}. There is a wide ranging review of this topic in the paper of \citet{LB96} in memory of Chandrasekhar.

In the present paper we review the developed approximate model for consideration of dynamics of collapsing ellipsoidal figures. Present review is based on the papers of \cite{BK2004}, \citet{BKTs2005}, \citet{BKTs2008}. We derive and solve equations of the dynamical behavior in a
simple model of a compressible uniformly rotating three-axis ellipsoid. The ordinary differential equations for ellipsoid semi-axes evolution in time are obtained with help of variation of the Lagrange function of ellipsoid. In our model a motion along three axes takes place in
the gravitational field of a uniform density ellipsoid, with
account of rotation, represented by centrifugial forces, and the isotropic pressure, represented in an approximate
non-differential way.

Our model can be used for investigation of star dynamics, and also for the first stages of formation of large scale structure consisted from dark matter, with appropriate approximations. Starting from section \ref{section-intro1}, we apply our model for investigation of dynamic stability of non-spherical stars.
We
show, that deviation from the spherical symmetry in a non-rotating
star with zero angular momentum leads to formation of a regularly or chaotically oscillating
body, in which dynamical motion prevents the formation of the
singularity. The non-spherical star without dissipative processes never will
reach a singularity in the Newtonian dynamics. Therefore collapse to a singularity is
connected with a secular type of instability, even without
rotation. Starting from section \ref{section-intro2}, we consider approximate model for large scale structure formation. We  estimate emission of a very long gravitational waves during a collapse of the dark matter, and discuss a possibility of gravitational lensing, and polarisation of CMB by these waves.

Recently \cite{vand2011} models the galaxy as a heterogeneous ellipsoid with an arbitrary stratification of the density distribution, and represent the mean motions of the stars in terms of a velocity field that sustains that density distribution consistently with the equation of continuity. The virial equations are reduced to a closed system of equations of motion that govern the semi-axes of the characteristic ellipsoid. The semi-axes of an ellipsoid characterizing the size and shape of the galaxy are functions of the time, whereas the stratification of the density does not change. In the second paper \citep{vand2014} author describes an application of the model to an investigation of chaotic behaviour in axisymmetric pulsations of non-rotating, spherically symmetric configurations and of rotating, axisymmetric configurations. See also  papers of \cite{Rodrigues2014}, \cite{Sharif2013}, \cite{Sharif2014}.

\section{Dynamic stability of non-spherical stars}
\label{section-intro1}

Dynamic stability of spherical stars is determined by an average
adiabatic power
\begin{equation}
\gamma = \left. \frac{\partial\log P}{\partial\log \rho}\right|_S .
\end{equation}
For a density distribution
\begin{equation}
\rho=\rho_0 \, \varphi\left(\frac{m}{M}\right) ,
\end{equation}
the star in the newtonian gravity is stable against
dynamical collapse when \citep{ZN, BK1989}
\begin{equation}
\int_0^R{\left(\gamma-\frac{4}{3}\right)P\frac{dm}{\varphi(m/M)}}>0 \, .
\end{equation}
Here $\rho_0$ is a central density, $M$ is a stellar mass, $m$ is the mass inside a
Lagrangian radius $r$, so that $m=4\pi\int_0^r{\rho r^2 dr}$, $M=m(R)$, $R$ is a stellar radius, $\varphi\left(\frac{m}{M}\right)$
is the Emden function corresponding to $\gamma=4/3$.

 Let us consider equation of state $P=K\rho^{\gamma}$. Spherical star with such equation of state is stable against collapse if $\gamma > 4/3$
 and unstable at $\gamma < 4/3$, when the instability leads to a collapse to singularity. At $\gamma = 4/3$ there is only one equilibrium state
 for a given mass, at a unique value of the constant $K_0$. This state is neutral to the stability, and may suffer contraction to singularity
 at $K<K_0$, or expansion to infinity at $K>K_0$.

Collapse of a spherical, initially unstable star
may be stopped only by a stiffening of the equation of state, like
neutron star formation at late stages of evolution, or formation
of fully ionized stellar core with $\gamma=\frac{5}{3}$ at the
collapse of clouds during star formation. Without such stiffening a
spherical star in the newtonian theory would collapse into a point
with an infinite density (singularity).

In the presence of a rotation a star is becoming more dynamically
stable against collapse. Due to the more rapid increase of a
centrifugal force during contraction, in comparison with the
newtonian gravitational force, collapse of a rotating star will be
always stopped at finite density by centrifugal forces.

Here we
show, that deviation from the spherical symmetry in a non-rotating
star with zero angular momentum leads to a similar stabilization,
and non-spherical star without dissipative processes never will
reach a singularity. Therefore collapse to a singularity is
connected with a secular type of instability, even without
rotation.

We calculate a dynamical behavior of a non-spherical, non-rotating
star after its loss of a linear stability, and investigate
nonlinear stages of contraction. We use approximate system of
dynamic equations, describing 3 degrees of freedom of a uniform
self-gravitating compressible ellipsoidal body
\citep{BKTs2008}. We obtain that the development of instability
leads to the formation of a regularly or chaotically oscillating
body, in which dynamical motion prevents the formation of the
singularity. We find regions of chaotic and regular pulsations by
constructing a Poincar\'e diagram for different values of the
initial eccentricity and initial entropy. For simplicity we
restrict ourself by calculating only spheroidal figures with
$\gamma=\frac{4}{3}$, see also results for
$\gamma=\frac{6}{5}$ in \cite{BKTs2008}.

\section{Equations of motion for semi-axes of rotating 3-axis ellipsoid}

Here we derive equations of motion for general case of rotating 3-axis ellipsoid and consider cases $\gamma=5/3$ and $\gamma=4/3$.

Let us consider a compressible 3-axis ellipsoid with semi-axes $a \neq b \neq c$
\begin{equation}
\label{ellipsoid}\frac{x^2}{a^2}+\frac{y^2}{b^2}+\frac{z^2}{c^2}=1.
\end{equation}
We assume that the ellipsoid rotates uniformly with an angular velocity $\Omega$ around the axis $z$. Let us approximate the density of the matter $\rho$
in the ellipsoid as a uniform. During collapse and posterior time evolution the sizes of semi-axes $a$, $b$, $c$, angular velocity $\Omega$ and density $\rho$ are changed with time, but ellipsoid keeps ellipsoidal shape with the homogeneous distribution of density and uniform rotation.

The constant mass $m$ and total angular momentum $M$ of the uniform
ellipsoid are connected with uniform density, angular velocity and
semi-axes as ($V$ is the volume of the ellipsoid)
\begin{equation}
\label{mass-moment}m = \rho \, V = \frac{4\pi}{3} \, \rho \, abc
\, , \quad M = \rho\,\Omega\int\limits_V (x^2+y^2) \, dV =
\frac{m}{5} \, \Omega (a^2+b^2) \, .
\end{equation}
We assume a linear dependence of the velocities on the coordinates in
the rotating frame
\begin{equation}
\label{linear-dep} \upsilon_x =\frac{\dot{a}x}{a} \: , \quad
\upsilon_y =\frac{\dot{b}y}{b} \: , \quad \upsilon_z
=\frac{\dot{c}z}{c} \, .
\end{equation}
If an homogeneous ellipsoidal body of inviscid fluid initially has velocities that are linear functions of position, then it will remain ellipsoidal and the internal velocities will remain linear functions of position for all time provided that the density remains spatially uniform (see, for example, \cite{LB96}).

In absence of any dissipation this
ellipsoid is a conservative system. To derive the equations of
motion (equations describing time evolutions of semi-axes) let write a Lagrange function of the ellipsoid which consists of kinetic energy $U_{kin}$ and potential energy $U_{pot}$ of ellipsoid:
\begin{equation}
\label{Lagr-funct0} L = U_{kin} - U_{pot}\, , \quad U_{pot} = U_g +
E_{th} + U_{rot} \: .
\end{equation}
The kinetic energy is written as (with using (\ref{linear-dep}))
\begin{equation}
\label{U-kin} U_{kin} = \frac{1}{2} \,
\rho\int\limits_V(\upsilon_x^2+\upsilon_y^2+\upsilon_z^2)\,dV
 = \frac{m}{10} \,
(\dot{a}^2+\dot{b}^2+\dot{c}^2) \: .
\end{equation}
The gravitational energy $U_g$ of the uniform ellipsoid is defined as
\citep{ll93}:
\begin{equation}
\label{grav-energy}
U_g=-\frac{3Gm^2}{10}\int\limits_0^{\infty}\frac{du}{\sqrt{(a^2+u)(b^2+u)(c^2+u)}}.
\end{equation}
In case of spheroids ($a=b \neq c$) this integral can be taken analytically, in case of ellipsoid it is expressed in elliptical integrals (see \cite{BK2004}, \cite{BKTs2005}).

The rotational energy is
\begin{equation}
\label{U-rot-omega} U_{rot} = \frac{1}{2}\,\rho\int \limits_V
v_{rot}^2 \,dV = \frac{1}{2}\,\rho\,\Omega^2 \int \limits_V
(x^2+y^2) \,dx\,dy\,dz = \frac{m}{10}\,\Omega^2 (a^2+b^2) \: .
\end{equation}
The angular velocity $\Omega$ is changed with time, and it is more convenient to express (\ref{U-rot-omega}) via $M$ which is constant in absence of dissipation. Taking into account the second expression in (\ref{mass-moment}),
we obtain the relation
\begin{equation}
\label{U-rot-M} U_{rot} = \frac{5}{2} \, \frac{M^2}{m(a^2+b^2)} \:
.
\end{equation}
Different form of thermal energy $E_{th}$ are used depending on problem under consideration.\\

\subsection{Rotating ellipsoid, case $\gamma=5/3$} \label{section-53}

First, we use the non-relativistic equation of state $P=K\rho^\gamma$ with polytropic index $\gamma = 5/3$. It corresponds to the case of a matter consisted of non-collisional non-relativistic dark matter particles.

The total thermal energy, with adiabatic index $n=1.5$ ($\gamma=5/3$), of non-relativistic dark matter
particles in the ellipsoid is $E_{th} \sim V^{-2/3} \sim (abc)^{-2/3}$, and the
relation between pressure $P$ and thermal energy $E_{th}$ is
$E_{th} = \frac{3}{2} P V$. Let us write $E_{th}$ using initial values $E_{th,in}$, $a_{in}$, $b_{in}$, $c_{in}$ as
\begin{equation}
\label{E-th}
E_{th} = \frac{E_{th,in} (a_{in}b_{in}c_{in})^{2/3}} {(abc)^{2/3}} = \frac
{\varepsilon} {(abc)^{2/3}} \, .
\end{equation}
We introduce here $\varepsilon$, which we will call as the entropy function.
The entropy function $\varepsilon$ is constant in the conservative
case, but it is variable in presence of a dissipation.

By variation of the Lagrange function we obtain the Lagrange equations
of motion:
\begin{equation} \label{a-53}
\ddot{a} = - \frac{3Gm}{2} \, a
\int\limits_0^{\infty}\frac{du}{(a^2+u)\Delta} \,\, +
\,\frac{10}{3m}\, \frac{1}{a} \,\frac {\varepsilon} {(abc)^{2/3}}
\,\,+ \frac{25 M^2}{m^2} \, \frac{a}{(a^2+b^2)^2} \, ,
\end{equation}
\begin{equation} \label{b-53}
\ddot{b} = - \frac{3Gm}{2} \, b
\int\limits_0^{\infty}\frac{du}{(b^2+u)\Delta} \,\,+
\,\frac{10}{3m}\, \frac{1}{b} \,\frac {\varepsilon} {(abc)^{2/3}}
\,\, + \frac{25 M^2}{m^2} \, \frac{b}{(a^2+b^2)^2} \, ,
\end{equation}
\begin{equation} \label{c-53}
\ddot{c} = - \frac{3Gm}{2} \, c
\int\limits_0^{\infty}\frac{du}{(c^2+u)\Delta} \,\,+
\,\frac{10}{3m}\, \frac{1}{c} \,\frac {\varepsilon} {(abc)^{2/3}}
\; , \; \; \Delta^2 = (a^2+u)(b^2+u)(c^2+u) \, .
\end{equation}
At the right-hand side of these equations there are forces per unit ellipsoid mass. The first term is the gravitational force directed to the center of ellipsoid, the second term is the force of thermal pressure directed outward, and the third term is the centrifugal force directed outward.

It is easy to check that equilibrium solution of these equations
are the Maclaurin sphe\-roid and the Jacobi ellipsoid \citep{cha}.\\

\subsection{Non-rotating ellipsoid, $\gamma=4/3$}

For investigation of dynamic stability of non-spherical bodies  we  consider the non-rotating compressible homogeneous ellipsoid with the equation of state $P=K\rho^\gamma$,
$\gamma =4/3$. The case $\gamma=5/3$ considered above is not
interesting for investigations of dynamic stability, because isentropic spherical star
with $\gamma=5/3$ always stops contraction, and never suffers
collapse to singularity.

A spherical star with $\gamma =4/3$
collapses to singularity at small enough $K$, and we show here, how
deviations from a spherical form prevent formation of any
singularity. For $\gamma = 4/3$, the thermal energy of the ellipsoid
is $E_{th} \sim V^{-1/3} \sim (abc)^{-1/3}$, and the value
\[
\varepsilon=E_{th}(abc)^{1/3} =3\left(\frac{3m}{4\pi}\right)^{1/3}K
\]
remains constant in time. A Lagrange function of the ellipsoid is
written as
\begin{equation}
\label{Lagr-funct} L = U_{kin} - U_{pot}\, , \quad U_{pot} = U_g +
E_{th} \: , \quad E_{th} = \frac {\varepsilon} {(abc)^{1/3}} \: .
\end{equation}

Equations of motion describing behavior of 3 semiaxes $(a,b,c)$ is
obtained from the Lagrange function (\ref{Lagr-funct}) in the form
\begin{equation} \label{a-43}
\ddot{a} = - \frac{3Gm}{2} \, a
\int\limits_0^{\infty}\frac{du}{(a^2+u)\Delta} \,\, +
\,\frac{5}{3m}\, \frac{1}{a} \,\frac {\varepsilon} {(abc)^{1/3}}
\, \, ,
\end{equation}
\begin{equation} \label{b-43}
\ddot{b} = - \frac{3Gm}{2} \, b
\int\limits_0^{\infty}\frac{du}{(b^2+u)\Delta} \,\,+
\,\frac{5}{3m}\, \frac{1}{b} \,\frac {\varepsilon} {(abc)^{1/3}}
\, \, ,
\end{equation}
\begin{equation} \label{c-43}
\ddot{c} = - \frac{3Gm}{2} \, c
\int\limits_0^{\infty}\frac{du}{(c^2+u)\Delta} \,\,+
\,\frac{5}{3m}\, \frac{1}{c} \,\frac {\varepsilon} {(abc)^{1/3}}
\; ,
\end{equation}
\[
\Delta^2 = (a^2+u)(b^2+u)(c^2+u) \, .
\]

\section{Dynamic stabilization of non-spherical bodies against unlimited collapse, numerical results}

To obtain a numerical solution of equations (\ref{a-43}), (\ref{b-43}), (\ref{c-43}) we write them in
non-dimensional variables. Let us introduce the variables
\[
\tilde{t} = \frac{t}{t_0}, \; \tilde{a} = \frac{a}{a_0}, \;
\tilde{b} = \frac{b}{a_0}, \; \tilde{c} = \frac{c}{a_0}, \;
\]
\[
\tilde{m} = \frac{m}{m_0}, \; \tilde{\rho} = \frac{\rho}{\rho_0},
\; \tilde{U} = \frac{U}{U_0}, \; \tilde{E}_{th} =
\frac{E_{th}}{U_0}, \; \tilde{\varepsilon} =
\frac{\varepsilon}{\varepsilon_0}.
\]
The scaling parameters $\, t_0, \; a_0, \; m_0, \; \rho_0, \; U_0,
\; \varepsilon_0$ are connected by the following relations
\begin{equation}
\label{scal-par} t_0^2 = \frac{a_0^3}{G m_0}, \, U_0 = \frac{G
m_0^2}{a_0}, \, \rho_0 = \frac{m_0}{a_0^3}, \, \varepsilon_0 = U_0
a_0.
\end{equation}
System of non-dimensional equations:
\begin{equation}
\label{eq28}
 \ddot{a} = - \frac{3m}{2} \, a
\int\limits_0^{\infty}\frac{du}{(a^2+u)\Delta} \,\, +
\,\frac{5}{3m}\, \frac{1}{a} \,\frac {\varepsilon} {(abc)^{1/3}}
\, \, ,
\end{equation}
\begin{equation}
\label{eq29} \ddot{b} = - \frac{3m}{2} \, b
\int\limits_0^{\infty}\frac{du}{(b^2+u)\Delta} \,\,+
\,\frac{5}{3m}\, \frac{1}{b} \,\frac {\varepsilon} {(abc)^{1/3}}
\, \, ,
\end{equation}
\begin{equation}
\label{eq30} \ddot{c} = - \frac{3m}{2} \, c
\int\limits_0^{\infty}\frac{du}{(c^2+u)\Delta} \,\,+
\,\frac{5}{3m}\, \frac{1}{c} \,\frac {\varepsilon} {(abc)^{1/3}}
\; ,
\end{equation}
\[
\Delta^2 = (a^2+u)(b^2+u)(c^2+u) \, .
\]

In equations (\ref{eq28})-(\ref{eq30}) only non-dimensional
variables are used, and "tilde" sign is omitted for simplicity in
this section. The non-dimensional Hamiltonian (or non-dimensional
total energy) is:
\[
H = U_{kin} + U_g + E_{th} = \frac{m}{10} \,
(\dot{a}^2+\dot{b}^2+\dot{c}^2) -
\]
\begin{equation}
\label{ham} -
\frac{3m^2}{10}\int\limits_0^{\infty}\frac{du}{\sqrt{(a^2+u)(b^2+u)(c^2+u)}}
+ \frac {\varepsilon} {(abc)^{1/3}} \, .
\end{equation}
In case of the sphere ($a=b=c$, $\dot{a}=\dot{b}=\dot{c}$) the
non-dimensional Hamiltonian and non-dimensional equations of
motion reduce to:
\begin{equation} \label{sphere-H}
H = \frac{3}{10} \,m \dot{a}^2 -\frac{3}{5a} \left( m^2 -
\frac{5}{3} \, {\varepsilon} \right) ,
\end{equation}
\begin{equation}
\label{sphere1-H} \ddot{a} = - \frac{1}{m a^2} \left( m^2 -
\frac{5}{3} \, \varepsilon \right) .
\end{equation}
As follows from (\ref{sphere-H}), (\ref{sphere1-H}) for the given
mass there is only one equilibrium value of $\varepsilon$
\begin{equation}
\label{sphere2-H}
 \varepsilon_{eq}=\frac{3m^2}{5},
\end{equation}
at which the spherical star has zero total energy, and it may have
an arbitrary radius. For smaller $\varepsilon<\varepsilon_{eq}$
the sphere should contract to singularity, and for
$\varepsilon>\varepsilon_{eq}$ there will be a total disruption of
the star with an expansion to infinity. We solve here numerically
the equations of motion for a spheroid with $a=b \neq c$, which,
using (\ref{eq28})-(\ref{eq30}), are written for the oblate
spheroid with $k=c/a<1$ as

\begin{equation}
\label{eq7} \ddot{a}=\frac{3}{2}\frac{m}{a^2(1-k^2)}
\biggl[k-\frac{\arccos{k}}{\sqrt{1-k^2}}\biggr]+\, \frac{5}{3m}\,
\frac{1}{a} \,\frac {\varepsilon} {(a^2c)^{1/3}},
\end{equation}

\begin{equation}
\label{eq8} \ddot{c}=-3\frac{m}{a^2(1-k^2)}
\biggl[1-\frac{k\arccos{k}}{\sqrt{1-k^2}}\biggr]+ \frac{5}{3m}\,
\frac{1}{c} \,\frac {\varepsilon} {(a^2c)^{1/3}};
\end{equation}
and for the prolate spheroid $k=c/a>1$ as

\begin{equation}
\label{eq7p} \ddot{a}=-\frac{3}{2}\frac{m}{a^2(k^2-1)}
\biggl[k-\frac{\cosh^{-1}{k}}{\sqrt{k^2-1}}\biggr]+ \frac{5}{3m}\,
\frac{1}{a} \,\frac {\varepsilon} {(a^2c)^{1/3}},
\end{equation}

\begin{equation}
\label{eq8p} \ddot{c}=3\frac{m}{a^2(k^2-1)}
\biggl[1-\frac{k\cosh^{-1}{k}}{\sqrt{k^2-1}}\biggr]+
\frac{5}{3m}\, \frac{1}{c} \,\frac {\varepsilon} {(a^2c)^{1/3}}.
\end{equation}
It is convenient to introduce variables

\begin{equation}
\label{eqvar}
 \varepsilon_*=
\frac{5}{3}\frac{\varepsilon}{m^2},\,\,\, t_*=t\sqrt m.
\end{equation}
In these variables the equations (\ref{eq7})-(\ref{eq8p}),
(\ref{sphere1-H}) are written as (omitting subscript "*")

\begin{equation}
\label{equ7} \ddot{a}=\frac{3}{2a^2(1-k^2)}
\biggl[k-\frac{\arccos{k}}{\sqrt{1-k^2}}\biggr]+\,
 \frac{1}{a} \,\frac {\varepsilon} {(a^2c)^{1/3}},
\end{equation}

\begin{equation}
\label{equ8} \ddot{c}=-\frac{3}{a^2(1-k^2)}
\biggl[1-\frac{k\arccos{k}}{\sqrt{1-k^2}}\biggr]+
 \frac{1}{c} \,\frac {\varepsilon} {(a^2c)^{1/3}}
\end{equation}
 for the oblate spheroid $k=c/a<1$,

\begin{equation}
\label{equ7p} \ddot{a}=-\frac{3}{2a^2(k^2-1)}
\biggl[k-\frac{\cosh^{-1}{k}}{\sqrt{k^2-1}}\biggr]+
 \frac{1}{a} \,\frac {\varepsilon} {(a^2c)^{1/3}},
\end{equation}

\begin{equation}
\label{equ8p} \ddot{c}=\frac{3}{a^2(k^2-1)}
\biggl[1-\frac{k\cosh^{-1}{k}}{\sqrt{k^2-1}}\biggr]+
 \frac{1}{c} \,\frac {\varepsilon} {(a^2c)^{1/3}}
\end{equation}
for the prolate spheroid $k=c/a>1$, and

\begin{equation}
\label{sphere3-H} \ddot{a} = - \frac{1 -  \varepsilon}{a^2}
\end{equation}
for the sphere, where the equilibrium corresponds to
$\varepsilon_{eq}=1$. Near the spherical shape we should use
expansions around $k=1$, what leads to equations of motion valid for
both oblate and prolate cases

\[
\ddot{a} = - \frac{1 -  \varepsilon}{a^2}+
\left(\frac{\varepsilon}{3}+\frac{3}{5}\right)\frac{1-k}{a^2},
\]
\begin{equation}
\label{sphere4-H} \ddot{c} = - \frac{1 -  \varepsilon}{a^2}+
\left(\frac{4\varepsilon}{3}-\frac{4}{5}\right)\frac{1-k}{a^2}.
\end{equation}
In these variables the total energy is written as

\begin{equation}
\label{eqvar1} H_*=\frac{H}{m^2},
\end{equation}
and omitting "*" we have

\[
H= \frac{\dot{a}^2}{5}+\frac{\dot{c}^2}{10}
-\frac{3}{5a}\frac{\arccos{k}}{\sqrt{1-k^2}} + \frac{3}{5}\frac
{\varepsilon} {(a^2c)^{1/3}},\quad {\rm (oblate)}
\]

\[
H= \frac{\dot{a}^2}{5}+\frac{\dot{c}^2}{10}
-\frac{3}{5a}\frac{\cosh^{-1}{k}}{\sqrt{k^2-1}} + \frac{3}{5}\frac
{\varepsilon} {(a^2c)^{1/3}}, \quad {\rm (prolate)}
\]
\begin{equation}
H = \frac{3\dot{a}^2}{10} -\frac{3}{5a} (1- {\varepsilon}), \quad
{\rm (sphere)} \label{ham1}
\end{equation}
\[
H= \frac{\dot{a}^2}{5}+\frac{\dot{c}^2}{10}
-\frac{3}{5a}\left(1+\frac{\delta}{3}+\frac{2\delta^2}{15}\right)
+\frac{3\varepsilon}{5a}\left(1+\frac{\delta}{3}+\frac{2\delta^2}{9}\right),\,\,\,
\]
\[
\delta=1-k,\,\,\,{\rm (around\,\,\, the\,\,\,
sphere)},\,\,|\delta|\ll 1.
\]

Solution of the system of equations (\ref{eq7})-(\ref{eq8p}) was
performed for initial conditions at $t=0$: $\dot c_0=0$, different
values of initial $a_0,\,\, \dot a_0,\,\,k_0$, and different
values of the constant parameter $\varepsilon$.
The results of our numerical calculations are

(i) sphere ($a=b=c$):

\quad \quad $\varepsilon = 1$ -- total energy equals to zero ($H = 0$), radius is arbitrary,

\quad \quad $\varepsilon < 1$ -- spherical star collapses to a singularity,

\quad \quad $\varepsilon > 1$ -- disruption of star with expansion to infinity;

\begin{figure}
\centerline{\hbox{\includegraphics[width=0.9\textwidth]{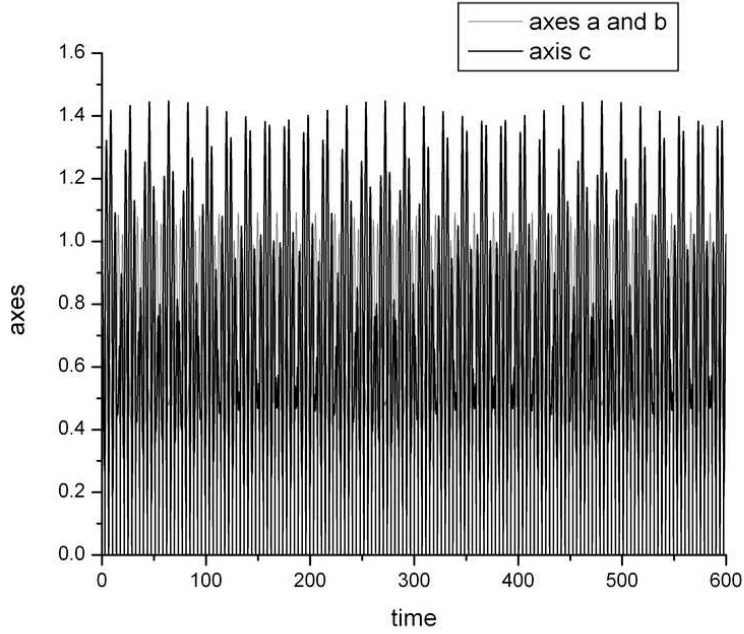}}}
\caption{Example of regular motion of spheroid with $\gamma=4/3$,
$H = - 1/5$, $\varepsilon=2/3$. This motion corresponds to full
line on the Poincar\'{e} map in Fig.\ref{fig-map1}.}
\label{fig-regular}
\end{figure}

\begin{figure}
\centerline{\hbox{\includegraphics[width=0.9\textwidth]{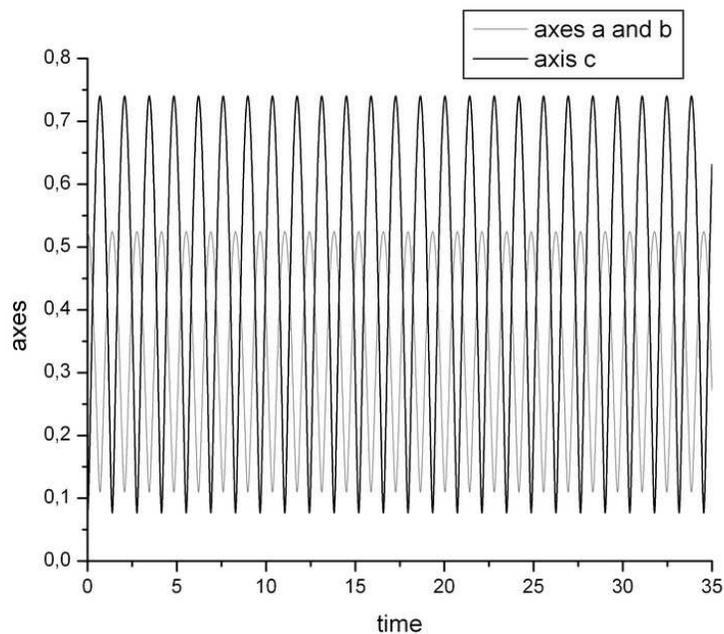}}}
\caption{Example of regular motion of spheroid with $\gamma=4/3$,
$H = - 1/5$, $\varepsilon=2/3$. This motion corresponds to the
point inside the regular region on the Poincar\'{e} map in Fig.\ref{fig-map1}.}
\label{fig-periodic}
\end{figure}

\begin{figure}
\centerline{\hbox{\includegraphics[width=0.9\textwidth]{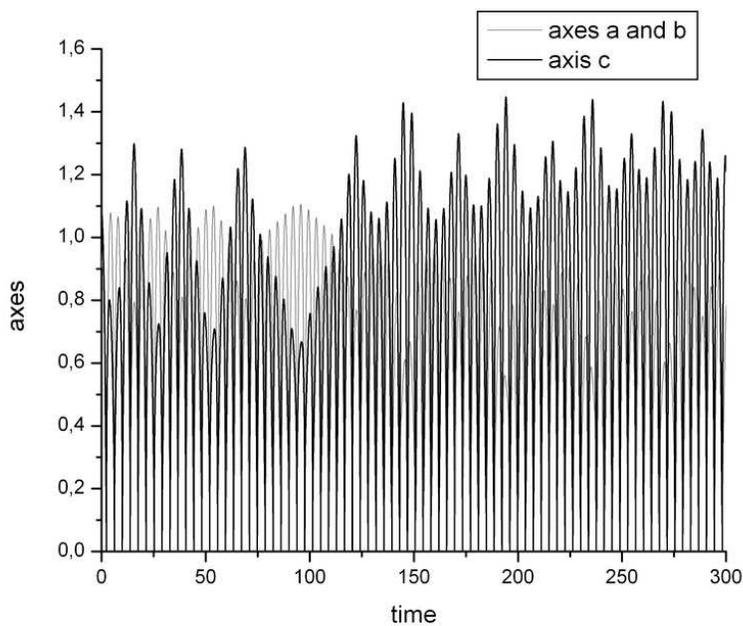}}}
\caption{Example of chaotic motion of spheroid with $\gamma=4/3$,
$H = - 1/5$, $\varepsilon=2/3$. This motion corresponds to gray
points on the Poincar\'{e} map in Fig.\ref{fig-map1zoom}.}
\label{fig-chaos}
\end{figure}

(ii) spheroid ($a=b \neq c$):



\quad \quad \quad at $\varepsilon  \ge 1$ the total energy is $H \ge 0$ -- disruption of star with expansion to infinity,

\quad \quad \quad at $\varepsilon < 1$ the total energy is $H < 0$ -- the oscillatory regime is established.
In the oscillatory regime the dynamical motion prevents the formation of the singularity. The type of oscillatory regime depends on initial conditions, and may be represented either by regular periodic oscillations, or by chaotic behavior. Examples of two types of
such oscillations are represented in Figs.\ref{fig-regular}--\ref{fig-periodic} (regular, periodic) and in Fig.\ref{fig-chaos} (chaotic). For rigorous separation between these
kinds of oscillations we use a method developed by Poincar\'{e}
\citep{LL83}. The same approach is used in recent papers of \cite{vand2011,vand2014}.

\section{Investigation of regular and chaotic oscillations via the Poincar\'{e} sections}

To investigate regular and chaotic dynamics we use the method of
Poincar\'{e} section \citep{LL83}, and obtain the Poincar\'{e} map
for different values of the total energy $H$. Let us consider a
spheroid with semi-axes $a=b \neq c$. This system has two degrees
of freedom. Therefore in this case the phase space is
four-dimensional: $a$, $\dot{a}$, $c$, $\dot{c}$. If we choose, at given $\varepsilon$, a
value of the Hamiltonian $H_0$, we fix a three-dimensional energy
surface $H(a,\dot{a},c,\dot{c}) = H_0$. During the integration of
the equations (\ref{eq7})-(\ref{eq8p}) which preserve the constant
$H$, we fix moments $t_i$, when $\dot{c}=0$. At these moments
there are only two independent values (i.g.  $a$ and $\dot{a}$),
because the value of $c$ is determined uniquely from the relation
for the hamiltonian at constant $H$. 

For the same values of $H$ and $\varepsilon$ we solve equations of
motion (\ref{eq7})-(\ref{eq8p}) at initial $\dot c=0$, and
different $a,\,\, \dot a$. For each integration we put the points
on the plane $(a, \dot{a})$ at the moments $t_i$. These points are
the intersection points of the trajectories on the
three-dimensional energy surface with a two-dimensional plane
$\dot{c}=0$, called a Poincar\'{e} section.

\begin{figure}
\centerline{\hbox{\includegraphics[width=0.9\textwidth]{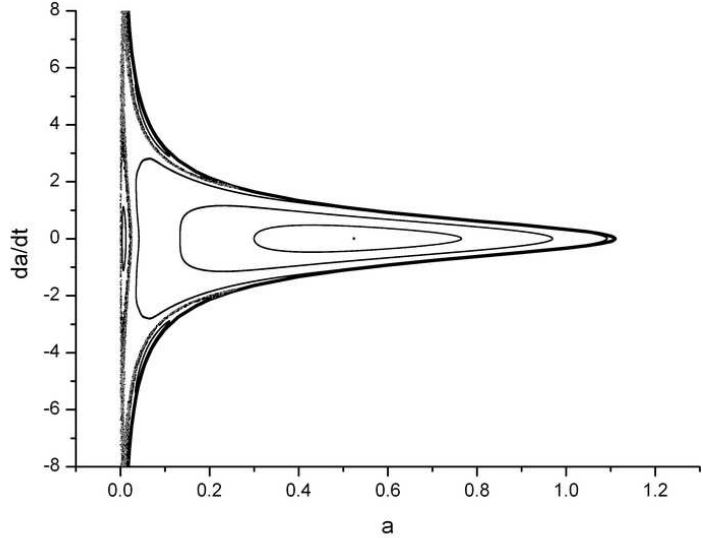}}}
\caption{The Poincar\'{e} map for five regular  and two chaotic
trajectories in case of $\gamma=4/3$, $H = - 1/5$,
$\varepsilon=2/3$. The $(a,\dot a)$ values are taken in the
minimum of c. Full black line is the bounding curve. The point
inside the regular region corresponds to coherent oscillations
with the same period for $a$ and $c$ values, represented in
Fig.2.}
\label{fig-map1}
\end{figure}

\begin{figure}
\centerline{\hbox{\includegraphics[width=0.9\textwidth]{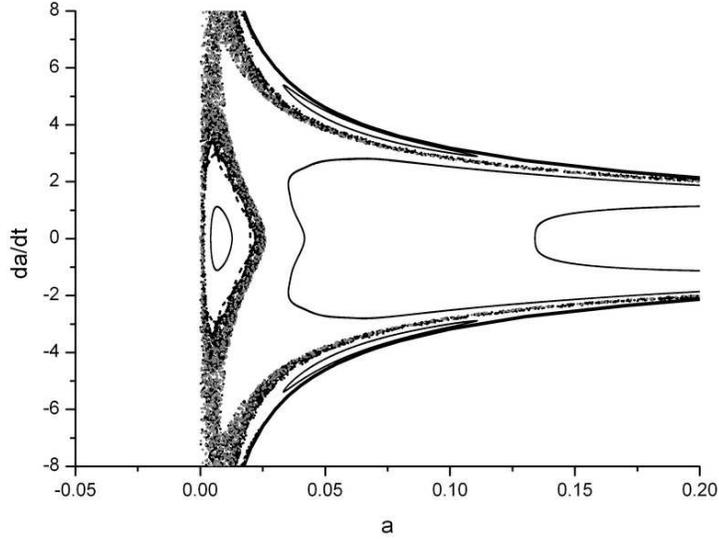}}}
\caption{Zoom of previous figure.}
\label{fig-map1zoom}
\end{figure}

\begin{figure}
\centerline{\hbox{\includegraphics[width=0.9\textwidth]{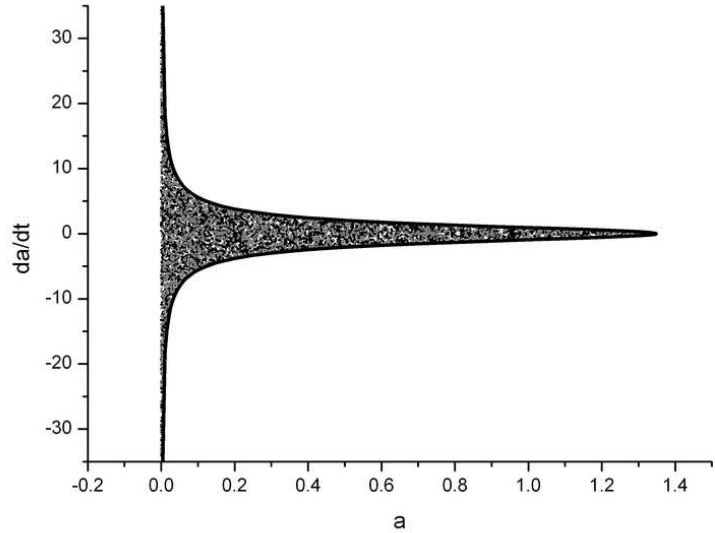}}}
\caption{The Poincar\'{e} map for two chaotic trajectories in case
of $\gamma=4/3$, $H = - 1/2$, $\varepsilon=1/6$. The $(a,\dot a)$
values are taken in the minimum of c. Full black line is the
bounding curve.}
\label{fig-map2}
\end{figure}

\begin{figure}
\centerline{\hbox{\includegraphics[width=0.9\textwidth]{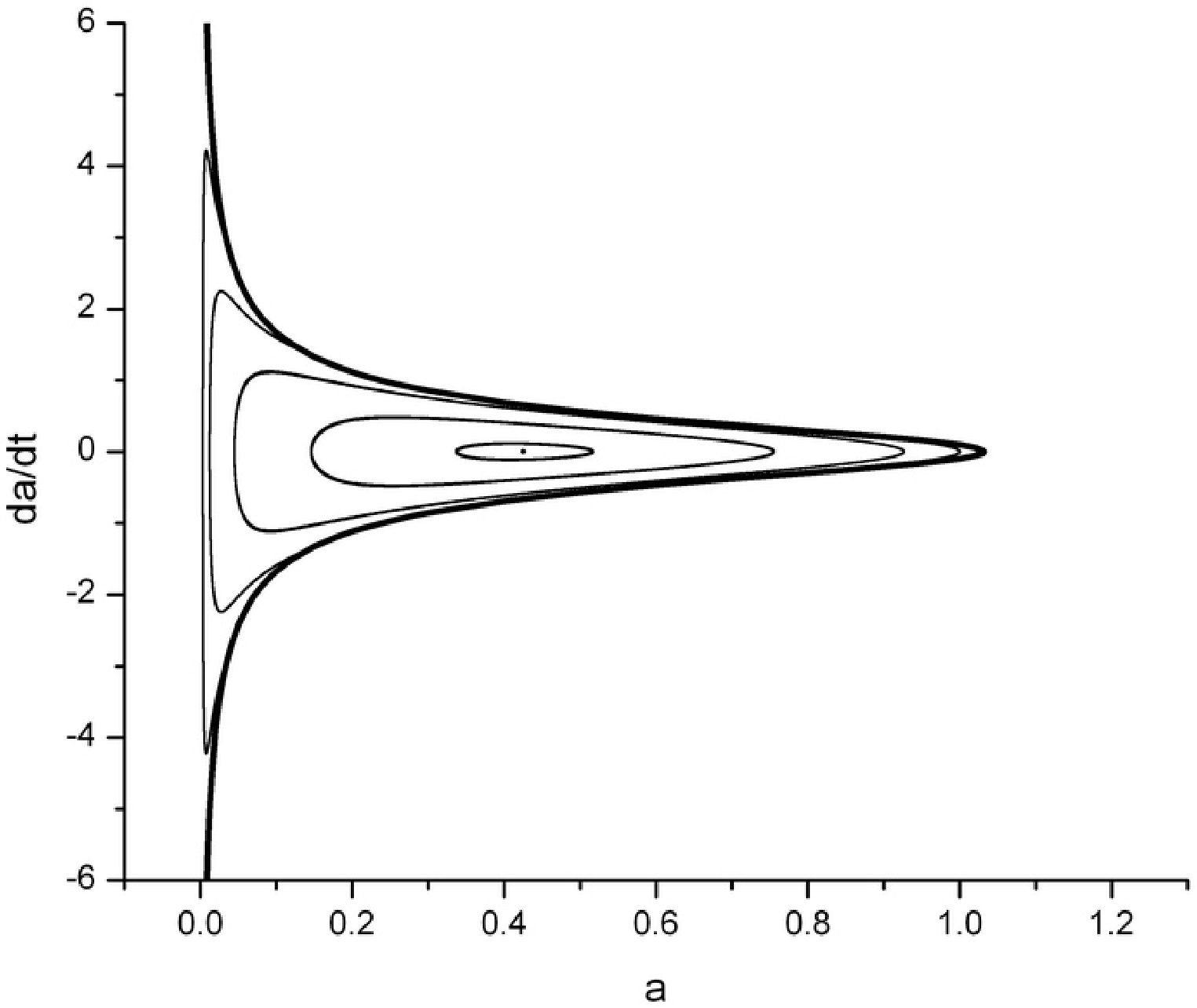}}}
\caption{The Poincar\'{e} map for six regular trajectories in the
case of $\gamma=4/3$, $H = - 3/50$, $\varepsilon=9/10$. The
$(a,\dot a)$ values are taken in the minimum of c. Full black line
is the bounding curve. The point inside the regular region
corresponds to coherent oscillations with the same period for $a$
and $c$ values, similar to those represented in Fig.\ref{fig-periodic}.}
\label{fig-map3}
\end{figure}

For each fixed combination of $\varepsilon, H$ we get the
Poincar\'{e} map, represented in Figs.\ref{fig-map1}, \ref{fig-map1zoom}, \ref{fig-map2}, \ref{fig-map3}. Condition $\dot{c}=0$
splits in two cases,  of a minimum and of a maximum of $c$. The
Poincar\'{e} maps are drawing separately, either for the minimum,
or for the maximum of $c$, and both maps lead to identical
results. The regular oscillations are represented by closed lines
on the Poincar\'{e} map, and chaotic behavior fills regions of
finite square with dots. These regions are separated from the
regions of the  regular oscillations by separatrix line.

\section{Dynamic stabilization of non-spherical bodies against unlimited collapse, discussion}

The main result following from our calculations is the indication
to a degenerate nature of formation of a singularity in unstable
newtonian self-gravitating gaseous bodies. Only pure spherical
models can collapse to singularity, but any kind of nonsphericity
leads to nonlinear stabilization of the collapse by a dynamic
motion, and formation of regularly or chaotically oscillating
body. This conclusion is valid for all unstable equations of
state, namely, for adiabatic with $\gamma < 4/3$. In addition to
the case with $\gamma =4/3$, we have calculated the dynamics of
the model with $\gamma=6/5$, and have obtained similar results.

Note, that region of chaotic behavior on the  Poincar\'{e} map is
gradually increasing for $\gamma=4/3$ with decreasing of the
entropy $\varepsilon$ and the total energy $H$. At
$\varepsilon=9/10$ and $H=-3/50$ we have found only regular
oscillations (Fig.\ref{fig-map3}), at $\varepsilon=2/3$ and $H=-1/5$ both kind
of oscillations are present (Figs.\ref{fig-map1}, \ref{fig-map1zoom}), and only chaotic behavior
is found at $\varepsilon=1/6$ and $H=-1/2$ (Fig.\ref{fig-map2}). We connect
\citep{BKTs2005} this chaotic behavior with development of
anisotropic instability, when radial velocities strongly exceed
the transversal ones \citep{ant, fp85}.

In reality a presence of dissipation leads to damping of these
oscillations, and to final collapse of nonrotating model, when
total energy of the body is negative. In the case of core-collapse
supernova the main dissipation is due to emission of neutrino. The
time of the neutrino losses is much larger than the characteristic
time of the collapse, so we may expect that the collapse leads to
formation of a neutron star where nonspherical modes are excited
and exist during several seconds after the collapse. In addition
to the damping due to neutrino emission the shock waves will be
generated, determining highly variable energy losses during the
oscillations. Besides viscosity and radiation which can damp the
ellipsoid-like motions and allow collapse, there is the
possibility that inertial interactions with higher order modes
that must be present in the real stratified bodies may cause an
inertial cascade and drain the energy from the second order modes
an a non-secular way that does not depend on dissipative
coefficients. It is very seductive to connect chaotic oscillations
with highly variable emission observed in the prompt gamma ray
emission of cosmic gamma ray bursts \citep{mg81}. A presence of
rotation and magnetic field strongly complicate the picture of the
core-collapse supernova explosion \citep{abkm05,mbka06}.

In this paper we consider in details only spheroidal bodies. In
reality the spheroid will become a triaxial ellipsoid during the
motion. In addition to spheroids we calculate many variants with
triaxial figures (see \citet{BKTs2005}). Qualitatively we
obtain the same results for ellipsoids: no singularity was reached
for any $\varepsilon > 0$, and establishing oscillatory (regular or
chaotic) regime under negative total energy prevents the collapse.
However, in the case of the ellipsoid with semi-axes $a \neq b
\neq c$ we have a system with three degrees of freedom and
six-dimensional phase space. Therefore we could not carry out
rigorous investigation of regular and chaotic types of motion by
the constructing Poincar\'{e} map as it was done for spheroid with
two degrees of freedom, and restricted ourselfs by a description
of the spheroidal case.

In the frame of a general relativity dynamic stabilization against
collapse by nonlinear nonspherical oscillations cannot be
universal. When the size of the body approaches gravitational
radius no stabilization is possible at any $\gamma$. Nevertheless,
the nonlinear stabilization may happen at larger radii, so after
damping of the oscillations the star would collapse to the black
hole. Due to development of nonspherical oscillations there is a
possibility for emission of gravitational waves during the
collapse of nonrotating bodies with the intensity similar to
rotating bodies, or even larger.

\section{Zeldovich's pancakes and large scale structure of Universe} \label{section-intro2}

One of the most important problem in modern cosmology is a formation of large scale structure of the Universe. Observations of large scale structure of Universe demonstrate strong inhomogeneity in density distribution.
\citet{z70a, z70b} has investigated the fragmentation of a homogeneous medium under the action of gravitation. It has been shown that an arbitrary perturbation in medium with zero pressure leads to formation of strongly non-spherical structures (discs), and only pure spherical perturbations lead to collapse into the point. These discs are known now as 'Zeldovich's pancakes'.

Depending on the medium properties, two types of pancakes can be formed:

(i) During contraction of the barionic gas the collapse is stopped by the pressure, the shock wave is formed. Behind the shock front the cooling, fragmentation and formation of vortexes occur, what is necessary for formation of spiral galaxies with large angular momentum.

(ii) During collapse of a  non-collisional dark matter the pancake is also formed, but no shocks are forming at the stage of maximal contraction. Medium is non-collisional, so after initial the collapse, subsequent oscillations occur. The particles fly through each other many times, and mixing in the phase space occurs -- so called violent relaxation. The system reaches quasi-equilibrium state via violent relaxation \citep{ref2} and phase mixing during the most rapid dynamic stages . Established equilibrium state has a form, depending on the angular momentum of rotation, and sizes depending on the final values of energy and entropy.

Subsequent nonlinear evolution and interaction between pancakes lead to formation of a very complicated large scale structure of matter in the Universe observed in the sky. It was shown by numerical simulations in two-dimensional space by \cite{Doroshkevich1980}. The first three-dimensional simulation were performed by \cite{Shandarin1983}, demonstrating
the origin of filaments in the dark matter density field. For review see \cite{Shandarin1989}
where the developments of the original Zeldovich's idea were summarized. The development of perturbations at non-linear stage and structure formation in the medium with barionic gas and non-collisional cold dark matter are investigated by using numerical modelling by many groups (see, for example,  \cite{Doroshkevich1999}, \cite{Boily1999}, \cite{Sneth2001}, \cite{Boily2002} ). For numerical modelling N-body simulations are used, which are very time consuming.

The first stage of large scale structure formation is a collapse of non-collisional dark matter with formation of a pancake and subsequent relaxation with formation of quasi-equilibrium structures. This stage can be investigated using approximate model of large scale structure formation considered in works of \cite{BK2004} and \cite{BKTs2005}. Form of the collapsing matter have been approximated as rotating homogeneous ellipsoid with two \citep{BK2004} and three \citep{BKTs2005} degrees of freedom. Correct description of pressure
effects, attained by such approach, and addition of a relaxation
permit to get the dynamics of motion without any numerical
singularities. In our model a motion along three axes takes place in
the gravitational field of a uniform density ellipsoid, with
account of the isotropic pressure, represented in an approximate
non-differential way. The relaxation leads to a transformation of
the kinetic energy of the ordered motion into the kinetic energy
of the chaotic motion, and to increase of the effective pressure
and thermal energy. All losses are connected with runaway
particles. The collapse of the rotating three-axis ellipsoid is
approximated by a system of ordinary differential equations, where
the relaxation, and the losses of energy, mass and angular
momentum are taken into account phenomenologically \citep{BK2004}. The system is solved numerically for several parameters, characterizing the
configuration.

\section{Equations of motion with dissipation and numerical results}

Equations of motion of rotating dark matter ellipsoid without any dissipation are derived in subsection \ref{section-53}, see eq. (\ref{a-53}), (\ref{b-53}), (\ref{c-53}).

In reality there is relaxation in the collisionless system,
connected with phase mixing, which is called "violent
relaxa\-tion" \citep{ref2}. This relaxation leads to a
transformation of the kinetic energy of the ordered motion into
the kinetic energy of the chaotic motion and increases effective
pressure and thermal energy.

The main transport process is an effective bulk viscosity.
Therefore there is a drag force, which is described
phenomenologically by adding of the terms
\begin{equation}
- \frac{\dot{a}}{\tau_{rel}} \: , \quad -
\frac{\dot{b}}{\tau_{rel}} \: , \quad - \frac{\dot{c}}{\tau_{rel}}
\end{equation}
in the right-hand parts of equations of motion.

Dissipation leads to a heat production.
The rate of a heat production leading to the growth of the entropy of
the spheroid matter was found in \cite{BK2004} from equations of motion
(\ref{a-53}), (\ref{b-53}), (\ref{c-53}) with account of
(\ref{Lagr-funct0}), allowing the energy, mass and angular momentum losses.

Here we have scaled the relaxation time $\tau_{rel}$ by the Jeans
characteristic time
\begin{equation}
\tau_J = \frac{2 \pi}{\omega_J} = \frac{2 \pi}{\sqrt{4 \pi G\rho}} =
2 \pi \sqrt{\frac{abc}{3Gm}}
\end{equation}
with a constant value of $\alpha_{rel}$
\begin{equation}
\label{tau-rel} \tau_{rel} = 2 \pi \, \alpha_{rel}
\sqrt{\frac{abc}{3Gm}} \, .
\end{equation} 
The energy balance relation follows from the dynamic equations in the form
\begin{equation}
\label{eq19}
\frac{dU_{tot}}{dt}=\frac{\partial U_g}{\partial m}\frac{dm}{dt}+
\frac{\partial U_{rot}}{\partial m}\frac{dm}{dt}+\frac{\partial U_{rot}}
{\partial M}\frac{dM}{dt}+
\frac{\partial E_{th}}{\partial {\cal E}}\frac{d{\cal E}}{dt}-2\frac{U_{kin}}
{\tau_{rel}}-\frac{U_{kin}}{m}\frac{dm}{dt},
\end{equation}
$$U_{tot}=U_{kin}+U_g+U_{rot}+E_{th}. $$ The process of relaxation
is accompanied also by the energy, mass, and angular momentum
losses from the system. We suggest, that these losses take place
only in non-stationary phases, so these rates are proportional to
the kinetic energy $U_{kin}$, the characteristic times for mass,
angular momentum and energy losses $\tau_{ml}$,
$\tau_{Ml}$,$\tau_{el}$ are considerably greater then
$\tau_{rel}$.  The equations describing different losses may be
phenomenologically written as
\begin{equation}
\label{eq20}
\frac{dU_{tot}}{dt}=-\frac{U_{kin}}{\tau_{ml}}, \quad
\frac{dm}{dt}=\frac{mU_{kin}}{U_g\tau_{ml}},\quad U_g<0,\quad
\frac{M}{M_{in}}=\left(\frac{m}{m_{in}}\right)^{\frac{\alpha_{ml}}
{\alpha_{Ml}}},
\end{equation}
where $m_{in}$ and $M_{in}$ are the initial values of corresponding parameters.
Scaling all the characteristic times by the Jeans value, we have
$\tau_{el}=\alpha_{el}\tau_J, \quad
\quad \tau_{ml}=\alpha_{ml}\tau_J, \quad \tau_{Ml}=\alpha_{Ml}\tau_J,
$
with constant values of $\alpha_i\,\, (i=el,\, ml,\, Ml)$.
The function ${\cal E}$ determines the entropy of the matter in the spheroid,
so that, using (\ref{E-th}), we have
\begin{equation}
\label{eq25}
\frac{dE_{th}}{dt}+ P\frac{dV}{dt}=\frac{1}{(abc)^{2/3}}
\frac{d{\cal E}}{dt},\,\,
E_{th}=\frac{\cal E}{(abc)^{2/3}}, \,\, V=\frac{4\pi}{3}abc,\,\,
P=\frac{2}{3}\frac{E_{th}}{V}.
\end{equation}
Using (\ref{eq20}) in (\ref{eq19}) we obtain the equation for
the entropy function $\cal E$ in presence of different losses
in the form
\begin{equation}
\label{eq26}
\frac{d{\cal E}}{dt}=(abc)^{2/3}U_{kin}\biggl[
\left(\frac{2}{\tau_{rel}}-\frac{1}{\tau_{el}}-\frac{2}{\tau_{ml}}\right)-
\frac{U_{rot}}{U_g}\left(\frac{2}{\tau_{Ml}}-\frac{1}{\tau_{ml}}\right)+
\frac{U_{kin}}{U_g\tau_{ml}}
\biggr].
\end{equation}

The equations, describing approximately the dynamics of the
formation of a stationary dark matter object include equations of
motion (\ref{a-53}), (\ref{b-53}), (\ref{c-53}) after adding terms with
relaxation; energy equation (\ref{eq26}) with account of
(\ref{U-kin}), (\ref{grav-energy}), (\ref{U-rot-M}); and equations
(\ref{eq20}), describing the losses of the energy,
mass and angular momentum.

To obtain a numerical solution of equations we write them in
non-dimensional variables. Let us introduce the variables
\[
\tilde{t} = \frac{t}{t_0} \: , \; \tilde{a} = \frac{a}{a_0} \:
, \; \tilde{b} = \frac{b}{a_0} \: , \; \tilde{c} =
\frac{c}{a_0} \: , \; \tilde{m} = \frac{m}{m_0} \: , \;
\tilde{M} = \frac{M}{M_0} \: , \; \tilde{\rho} =
\frac{\rho}{\rho_0} \: , \;\]

$$ \tilde{U} = \frac{U}{U_0} \: ,
\; \tilde{E}_{th} = \frac{E_{th}}{U_0} \: , \;
\tilde{\varepsilon} = \frac{\varepsilon}{\varepsilon_0} \: , \;
\tilde{\tau_i} = \frac{\tau_i}{t_0} \: .
$$
The scaling parameters $\, t_0, \; a_0, \; m_0, \; M_0, \; \rho_0,
\; U_0, \; \Omega_0, \; \varepsilon_0$ are connected by the
following relations
\[
t_0^2 = \frac{a_0^3}{G m_0}, \, M_0^2 = G a_0 m_0^3, \, U_0 =
\frac{G m_0^2}{a_0}, \, \rho_0 = \frac{m_0}{a_0^3}, \, \Omega_0 =
\frac{M_0}{m_0 a_0^2}, \, \varepsilon_0 = U_0 a_0^2.
\]
$U_0$ is used for scaling of all kind of energies. Hereinafter the
'tilde' sign is omitted for simplicity. In non-dimensional
variables we have $m = \frac{4\pi}{3} \, \rho \, abc, \; M =
\frac{m}{5} \, \Omega (a^2+b^2), \; \tau_i = 2 \pi \, \alpha_i
\sqrt{\frac{abc}{3m}}$.

Taking into account violent relaxation, total energy, mass and
angular momentum losses, the dynamics of the system is described
by the following non-dimensional system of equations
\begin{equation}
\label{ddota}
  \ddot{a} = - \frac{\dot{a}}{m} \frac{dm}{dt} -
\frac{3m}{2} \, a \int\limits_0^{\infty}\frac{du}{(a^2+u)\Delta}\,
+ \,\frac{10}{3m}\, \frac{1}{a} \,\frac {\varepsilon}
{(abc)^{2/3}} \, + \frac{25 M^2}{m^2} \, \frac{a}{(a^2+b^2)^2} \,
- \,\frac{\dot{a}}{\tau_{rel}},
\end{equation}

\begin{equation}
\label{ddotb}
 \ddot{b} = - \frac{\dot{b}}{m} \frac{dm}{dt} - \frac{3m}{2} \, b
\int\limits_0^{\infty}\frac{du}{(b^2+u)\Delta} \,\,+
\,\frac{10}{3m}\, \frac{1}{b} \,\frac {\varepsilon} {(abc)^{2/3}}
\,\, + \frac{25 M^2}{m^2} \, \frac{b}{(a^2+b^2)^2} -
\,\frac{\dot{b}}{\tau_{rel}} \, ,
\end{equation}

\begin{equation}
\label{ddotc}
 \ddot{c} = - \frac{\dot{c}}{m} \frac{dm}{dt}
 - \frac{3m}{2} \, c
\int\limits_0^{\infty}\frac{du}{(c^2+u)\Delta} \,\,+
\,\frac{10}{3m}\, \frac{1}{c} \,\frac {\varepsilon} {(abc)^{2/3}}
- \,\frac{\dot{c}}{\tau_{rel}} \, ,
\end{equation}

\begin{equation}
\label{ddote}
 \dot{\varepsilon} = (abc)^{2/3}\, U_{kin} \left[\left(
\frac{2}{\tau_{rel}} - \frac{1}{\tau_{el}} - \frac{2}{\tau_{ml}}
\right) - \frac{U_{rot}}{U_g} \left(\frac{2}{\tau_{Ml}} -
\frac{1}{\tau_{ml}} \right) + \frac{U_{kin}}{U_g \,
\tau_{ml}}\right] ,
\end{equation}

\begin{equation}
\label{ddotm}
 \dot{m} = - \frac{1}{3\tau_{ml}} \,
(\dot{a}^2+\dot{b}^2+\dot{c}^2) /
\left(\int\limits_0^{\infty}\frac{du}{\Delta}\right) , \; \;
\frac{M}{M_{in}} =
\left(\frac{m}{m_{in}}\right)^{\frac{\tau_{ml}}{\tau_{Ml}}} \; ,
\; \; \Delta^2 = (a^2+u)(b^2+u)(c^2+u) \, .
\end{equation}
This system is solved numerically for several initial parameters.
The case of spheroid ($a = b \neq c$) is consi\-dered in the paper of
\citet{BK2004}, the case of ellipsoid -- in the paper of \citet{BKTs2005}. For
investigation of the dynamical behavior we start the simulation from
spherical body of unit mass ($m_{in} = 1$), zero or small entropy
$\varepsilon \ll 1$. We also specify the parameters,
characterizing different dissipations. In all calculations with
relaxation we use the following values: $\alpha_{rel} = 3, \;
\alpha_{el} = \alpha_{ml} =  \alpha_{Ml} = 15$. We set non-zero
initial velocities. Because of the rotation around $c$-axis, the initial
sphere transforms to spheroid during the collapse, if initial
parameters for axes $a$ and $b$ are exactly equal ($a_{in} = b_{in}, \:
\dot{a}_{in} = \dot{b}_{in}$). Note that during the motion spheroids may be
not only oblate but prolate too \citep{BK2004}. To study an
appearance of 3-axis figures the initial parameters for axes $a$
and $b$ had been slightly different in all variants of
calculations.

\begin{figure}
\includegraphics[width=0.9\textwidth]{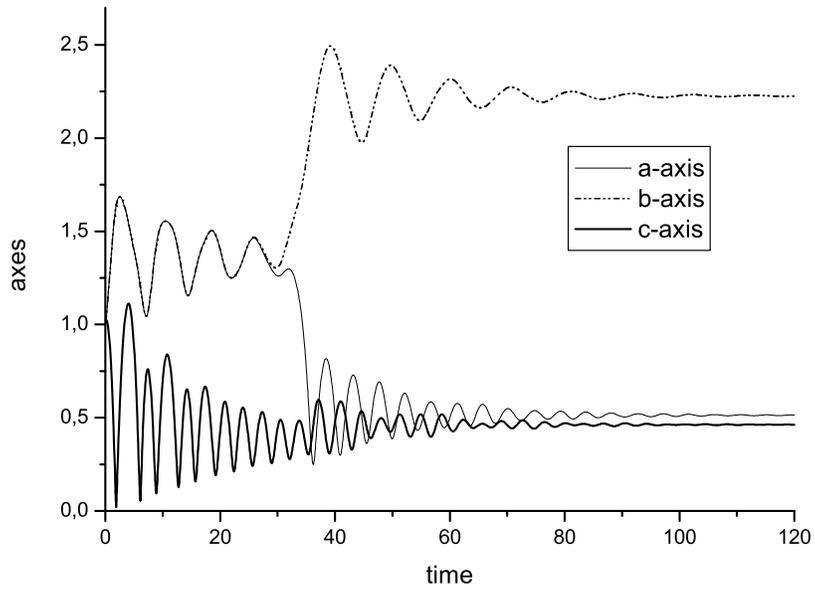}
\caption{Development of an instability at large angular momentum,
and formation of a stationary triaxial figure.}
\label{fig-05}
\end{figure}

\begin{figure}
\includegraphics[width=0.9\textwidth]{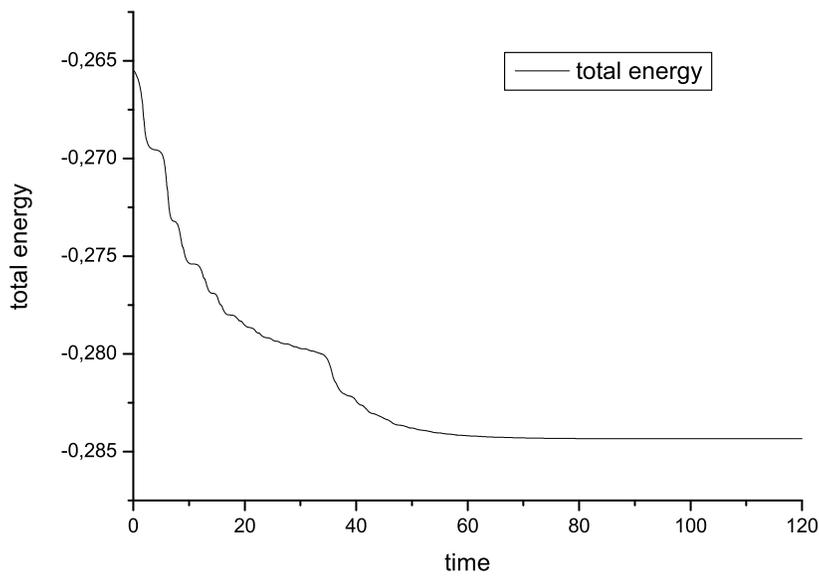}
\caption{The total energy evolution for the case considered on Fig.\ref{fig-05}.}
\label{fig-05energy}
\end{figure}

\begin{figure}
\includegraphics[width=0.9\textwidth]{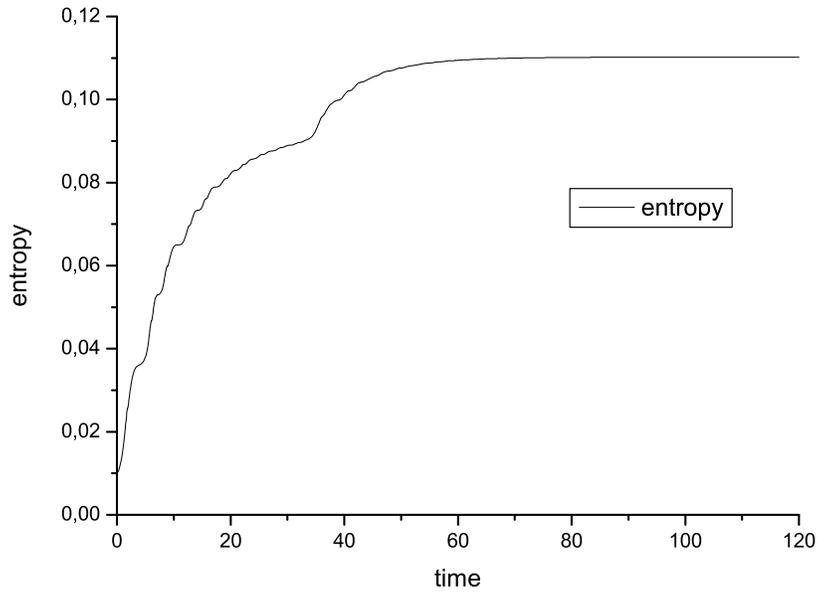}
\caption{The entropy function evolution for the case considered on Fig.\ref{fig-05}.}
\label{fig-05entr}
\end{figure}

\begin{figure}
\includegraphics[width=0.9\textwidth]{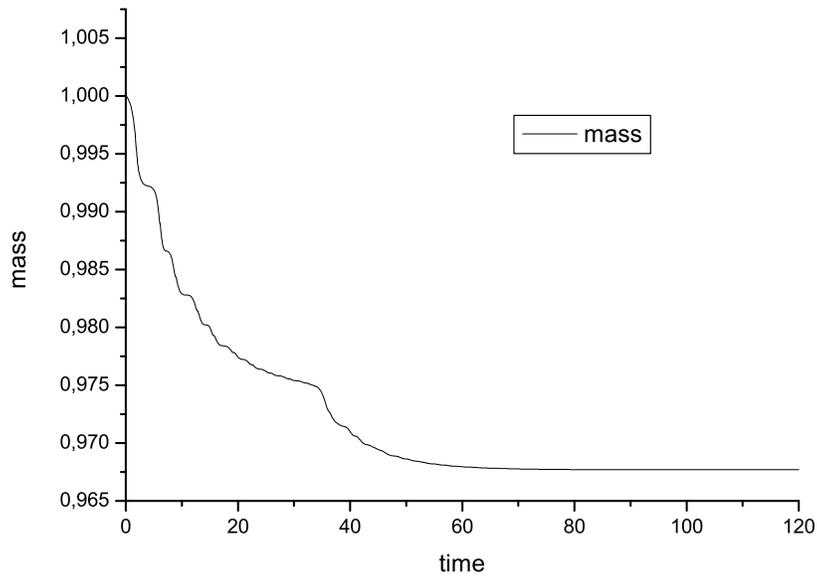}
\caption{The mass evolution for the case considered on Fig.\ref{fig-05}.}
\label{fig-05mass}
\end{figure}

In the first variant (Fig.\ref{fig-05}) there is a large initial angular
momentum $M_{in} = 0.5$. The field of velocities is slightly
perturbed by increasing $\dot{b}$ in comparison with $\dot{a}$.
Initially we observe the collapse and the formation of the
pancake. During the motion the difference $(a-b)$ increases due to
development of a secular instability, and we obtain the
transformation of the Maclaurin sphe\-roid into the Jacobi
el\-lip\-soid in the dynamics. Because of the relaxation, the
oscillating motion damps, and the configuration reaches the
equilibrium state of the 3-axis ellipsoid. The corresponding
behaviors of total energy $U_{tot}$, entropy function
$\varepsilon$ and mass $m$ are represented in Figs \ref{fig-05energy}, \ref{fig-05entr}, \ref{fig-05mass}. We see
that the main changes of these parameters take place at the stages
of first few oscillations: the entropy function $\varepsilon$
increases, the mass $m$ and the total energy $U_{tot}$ decreases.
Then these parameters reaches the equilibrium values.

\begin{figure}
\includegraphics[width=0.9\textwidth]{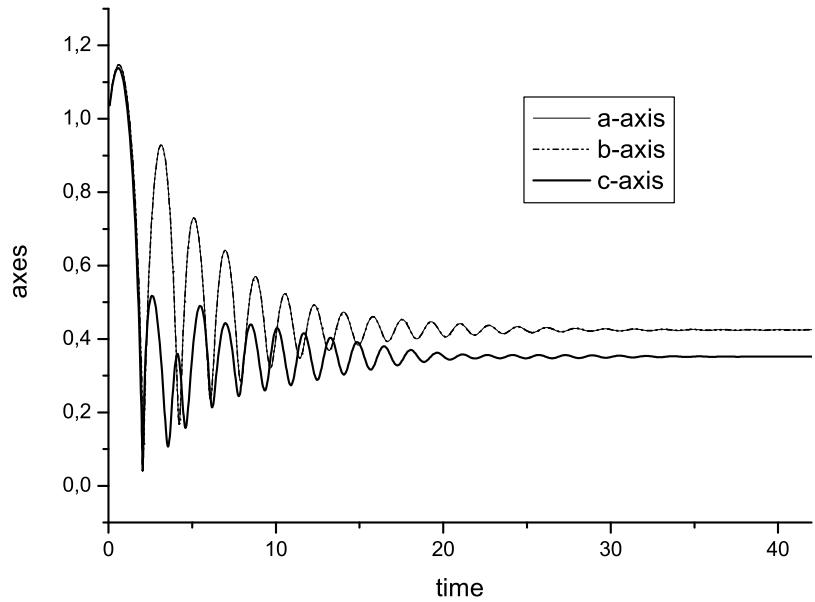}
\caption{The relaxation to the spheroid at small angular momentum.}
\label{fig-01}
\end{figure}

\begin{figure}
\includegraphics[width=0.9\textwidth]{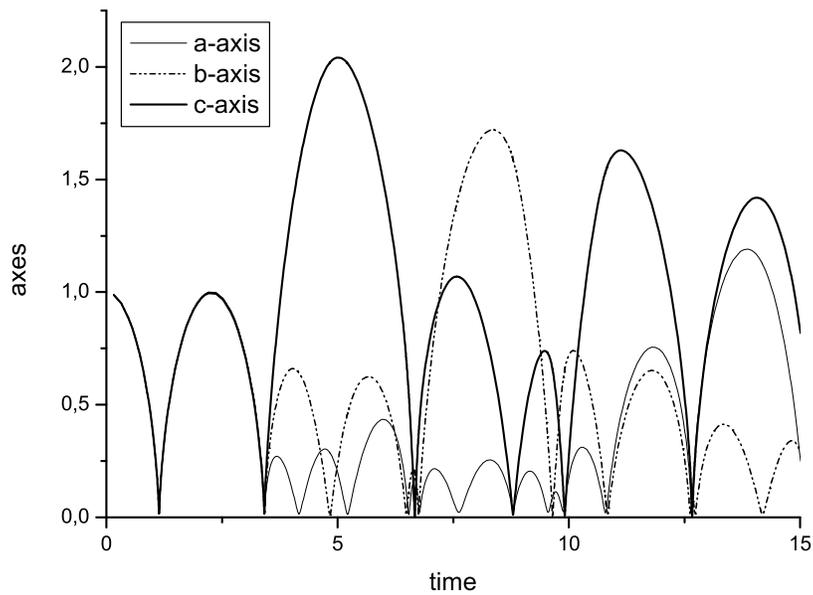}
\caption{Radial instability development in the non-rotating body,
without dissipation.}
\label{fig-radneust}
\end{figure}

\begin{figure}
\includegraphics[width=0.9\textwidth]{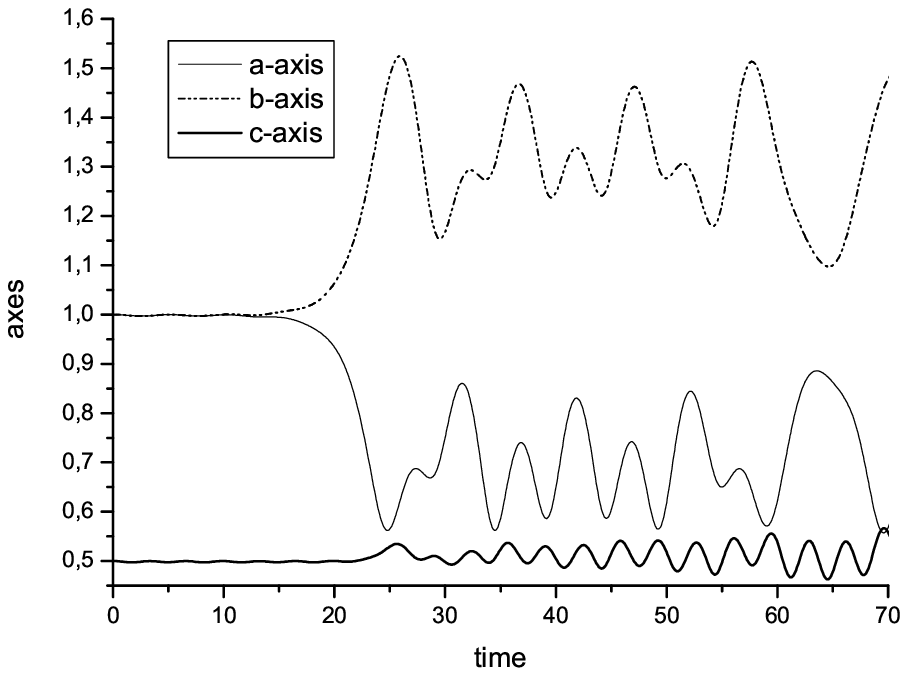}
\caption{Development of the secular instability in a rapidly
rotating body, without dissipation.}
\label{fig-secul}
\end{figure}

In the second variant of calculations (Fig.\ref{fig-01}) we set a small
initial angular momentum $M_{in}=0.1$. After the collapse, there
are temporary oblate and prolate spheroid appearance during
relaxation. There is no secular instability, and the system
rea\-ches the equilibrium oblate spheroid of Maclaurin. Here we have an additional
instability, there is increasing of the difference $(a-b)$ at
early stages of motion. However this difference remains too small
for visualizing in Figure. This is the instability characteristic
to the system with purely radial trajectories \citep{ant, fp85}.
This instability takes place in all variants of calculations, but
in most cases it is very small and always disappears during the
relaxation and formation of the equilibrium figure. In the cases
of large angular momentum this instability does not reveal because
the difference $(a-b)$ quickly increases as a result of the secular
instability.

For the illustration of the radial instability we consider initial
sphere without rotation ($M_{in} = 0$) and without any relaxation
process.  Only three equations (\ref{ddota})--(\ref{ddotc}) with
$\tau_{rel}=\infty$ have been used, at constant $m=1$, $M=0$ and
$\varepsilon$.  We have set small initial differences between axes
$a$, $b$ and $c$, and have obtained an unstable behavior evidently
visible in Fig.\ref{fig-radneust}.

For numerical investigation of the secular instability we consider
equations of motion without dissipation, and take the equilibrium
spheroids with small perturbation as the initial configuration. It
is easy to obtain equilibrium parameters of the spheroid from
equations of motion with zero accelerations. The initial
equilibrium configuration of the Maclaurin spheroid was slightly
perturbed by increasing $b$. At
small angular momentum we obtain the conservation of the initial
configuration. There are small oscillations around the equilibrium
spheroid, during which the difference $(a-b)$ changes sign
periodically. At large angular momentum we obtain the development
of the secular instability, and, after several low-amplitude
oscillations, the Maclaurin sphe\-roid,  is transformed into the
Jacobi el\-lip\-soid (see Fig.\ref{fig-secul} where the variant with large
angular momentum is represented). Because of absence of
relaxation, the system does not reach equilibrium configuration
and oscillations continue. If we include relaxation in this
variant, we should observe the final equilibrium state of
ellipsoid. Our calculations allow to find approximately
numerically the point of the onset of the secular instability of
the compressible Maclaurin sphe\-roid.

\section{Equilibrium configurations and stability}

Equilibrium of the uniformly rotating figure (spheroid or
ellipsoid) is found from the equations
(\ref{ddota})--(\ref{ddotc}) with zero time derivatives:

\begin{equation}
\label{a}
  0 = - \frac{3m}{2} \, a
\int\limits_0^{\infty}\frac{du}{(a^2+u)\Delta} \,\, +
\,\frac{10}{3m}\, \frac{1}{a} \,\frac {\varepsilon} {(abc)^{2/3}}
\,\,+ \frac{25 M^2}{m^2} \, \frac{a}{(a^2+b^2)^2} \, ,
\end{equation}
\begin{equation}
\label{b}
 0 = - \frac{3m}{2} \, b
\int\limits_0^{\infty}\frac{du}{(b^2+u)\Delta} \,\,+
\,\frac{10}{3m}\, \frac{1}{b} \,\frac {\varepsilon} {(abc)^{2/3}}
\,\, + \frac{25 M^2}{m^2} \, \frac{b}{(a^2+b^2)^2} \, ,
\end{equation}
\begin{equation}
 \label{c}
 0 = - \frac{3m}{2} \, c
\int\limits_0^{\infty}\frac{du}{(c^2+u)\Delta} \,\,+
\,\frac{10}{3m}\, \frac{1}{c} \,\frac {\varepsilon} {(abc)^{2/3}}.
\end{equation}
Introducing variables

\begin{equation}
 \label{var}
 x=\frac{u}{a^2},\,\,k=\frac{c}{a},\,\, k_1=\frac{b}{a},\,\,
 j=\frac{M}{m},\,\, \epsilon=\frac{\varepsilon}{m},
\end{equation}
and integrals

\begin{equation}
\label{vari}
I_1(k,k_1)=\int_0^{\infty}\frac{dx}{(x+1)^{3/2}\sqrt{(k_1^2+x)(k^2+x)}}
\, , \quad
I_2(k,k_1)=\int_0^{\infty}\frac{dx}{(x+k_1^2)^{3/2}\sqrt{(1+x)(k^2+x)}}
\, ,
\end{equation}
\[
I_3(k,k_1)=\int_0^{\infty}\frac{dx}{(x+k^2)^{3/2}\sqrt{(1+x)(k_1^2+x)}}
\, ,
\]
 we write the equations (\ref{a})-(\ref{c}), using (\ref{ddotm}) in the form

\begin{equation}
\label{a1}
  0 = - \frac{3m}{2} \, a \,I_1(k,k_1) +
\,\frac{10}{3}\, \frac {\epsilon} {(kk_1)^{2/3}} \,\,+ \frac{25
j^2}{(1+k_1^2)^2} \, ,
\end{equation}
\begin{equation}
\label{b1}
 0 = - \frac{3m}{2} \, ak_1 \,I_2(k,k_1) +
\,\frac{10}{3}\, \frac {\epsilon} {k_1(kk_1)^{2/3}} \,\,+ \frac{25
j^2\, k_1}{(1+k_1^2)^2} \, ,
\end{equation}
\begin{equation}
 \label{c1}
 0 = - \frac{3m}{2} \, ak \,I_3(k,k_1) +
\,\frac{10}{3}\, \frac {\epsilon} {k(kk_1)^{2/3}}.
\end{equation}
Excluding $m$, we obtain

\begin{equation}
\label{a2}
  0 = - \frac{1}{k^2}\frac{I_1(k,k_1)}{I_3(k,k_1)} +1
+ \frac{15 j^2}{2\epsilon}\frac{(kk_1)^{2/3}}{(1+k_1^2)^2} \, ,
\quad
 0 = - \frac{k_1^2}{k^2}\frac{I_2(k,k_1)}{I_3(k,k_1)} +1
+ k_1^2\frac{15 j^2}{2\epsilon}\frac{(kk_1)^{2/3}}{(1+k_1^2)^2}.
\end{equation}
At $a=b$, $k_1=1$ equations (\ref{a2}) are identical, and
determine the equilibrium of Maclaurin spheroid, see \cite{BK2004}.
For Jacobi ellipsoids with $a \neq b \neq c$, we obtain the
following relation between $k$ and $k_1$
\begin{equation}
\label{a3}
 F(k,k_1)=1-\frac{1}{k_1^2} + \frac{1}{k^2}\frac{I_2(k,k_1)}{I_3(k,k_1)}
 - \frac{1}{k^2}\frac{I_1(k,k_1)}{I_3(k,k_1)}=0 .
\end{equation}
This equation has a trivial solution $k_1=1$ at all $k$,
corresponding to the Maclairin spheroid. Let us find the
bifurcation point of the equation (\ref{a3}), at which nontrivial
solutions appear. While $k_1=1$ is always a root of the equation
(\ref{a3}), we may write $ F(k,k_1)=(k_1-1)f(k,k_1)$. Additional
root of equation $F(k,k_1)=0$ appears, when the root of the
equation $f(k,k_1)=0$ appears at $k_1=1$. The root of the zero
derivative equation
$F'_{k_1}(k,k_1)=f(k,k_1)+(k_1-1)f'_{k_1}(k,k_1)=0$ at $k_1=1$
coincides with the root of the equation $f(k,1)=0$ \footnote{We
are grateful to A.I.Neishtadt for useful discussion of this
point},  therefore the value of $k$ at the bifurcation point is
determined by the equation
\begin{equation}
\label{b3}
 \frac{\partial F(k,k_1)}{\partial k_1}|_{k_1=1}=0.
\end{equation}
Using (\ref{vari}) and (\ref{a3}), this equation is written in the
form
\begin{equation}
\label{b4}
 2k^2\,I_3(k,k_1)|_{k_1=1}+\frac{\partial I_2(k,k_1)}{\partial
 k_1}|_{k_1=1}-\frac{\partial I_1(k,k_1)}{\partial k_1}|_{k_1=1}=0
\end{equation}
At $k_1=1$ we have the analytic expressions
\begin{equation}
\label{b5}
 \frac{\partial I_1(k,k_1)}{\partial k_1}|_{k_1=1}=-I_0,\quad
 \frac{\partial I_2(k,k_1)}{\partial k_1}|_{k_1=1}=-3I_0,\quad
 I_0=\int_0^{\infty}\frac{dx}{(1+x)^3\sqrt{k^2+x}} \, ,
\end{equation}
so that the equation (\ref{b4}) is reduced to
\begin{equation}
\label{b6}
 k^2\,I_3=I_0,\quad {\rm where}\,\,\, I_3=I_3(k,1),\quad {\rm and}\,\,\, I_2=I_2(k,1)=
 I_1=I_1(k,1).
\end{equation}
Taking analytically the integrals $I_0,\,\, I_1=I_2$ and $I_3$ we
have

\begin{equation}
\label{b7}
 (1-k^2)I_0=\frac{3}{4}I_2-\frac{k}{2} \, ,\quad
 I_3=\frac{2}{k(1-k^2)}\biggl(1-\frac{k\arccos{k}}{\sqrt{1-k^2}}\biggr),
 \,
 \quad
\end{equation}
$$ I_1=I_2=-\frac{1}{k(1-k^2)}\biggl(k^2-\frac{k\arccos{k}}{\sqrt{1-k^2}}\biggr).
$$
Using (\ref{b7}) in (\ref{b6}) we obtain the equation

\begin{equation}
\label{b9} \frac{\arccos k}{\sqrt{1-k^2}} =
\frac{k(13-10k^2)}{3+8k^2-8k^4} \, ,
\end{equation}
which solution $k=0.582724$ determines the bifurcation point at
the sequence of the Maclaurin spheroids. For the uniform spheroid
the position of this point does not depend on the adiabatic index
of the matter.

Above we have obtained the bifurcation point on the equilibrium
curve of the Maclaurin spheroids using only the equilibrium
relations for the Jacobi ellipsoids. The usual way for
investigation of stability is to solve linearized equations of
motion and to find the eigenfrequencies.

Non-dimensional equations of motion without relaxation are
\begin{equation}
\label{DD-a} \ddot{a} = - \frac{3m}{2} \, a
\int\limits_0^{\infty}\frac{du}{(a^2+u)\Delta} \,\, +
\,\frac{10}{3m}\, \frac{1}{a} \,\frac {\varepsilon} {(abc)^{2/3}}
\,\,+ \frac{25 M^2}{m^2} \, \frac{a}{(a^2+b^2)^2} \, ,
\end{equation}
\begin{equation}
\label{DD-b} \ddot{b} = - \frac{3m}{2} \, b
\int\limits_0^{\infty}\frac{du}{(b^2+u)\Delta} \,\,+
\,\frac{10}{3m}\, \frac{1}{b} \,\frac {\varepsilon} {(abc)^{2/3}}
\,\, + \frac{25 M^2}{m^2} \, \frac{b}{(a^2+b^2)^2} \, ,
\end{equation}
\begin{equation}
\label{DD-c} \ddot{c} = - \frac{3m}{2} \, c
\int\limits_0^{\infty}\frac{du}{(c^2+u)\Delta} \,\,+
\,\frac{10}{3m}\, \frac{1}{c} \,\frac {\varepsilon} {(abc)^{2/3}}
\; , \; \; \Delta^2 = (a^2+u)(b^2+u)(c^2+u) \, .
\end{equation}
At given $a,b,c$ the equilibrium values of $\varepsilon$ è $M$ (the equilibrium configuration of ellipsoid) are obtained using formulae:
\begin{equation}
\label{ravn-eps} \varepsilon = \frac{9}{20} \, m^2 \, c^2 \,
(abc)^{2/3} \int\limits_0^{\infty}\frac{du}{(c^2+u)\Delta} \: ,
\end{equation}
\begin{equation}
\label{ravn-Mom} \frac{25 M^2}{m^2} \, \frac{a^2}{(a^2+b^2)^2} =
\frac{3m}{2} \, a^2 \int\limits_0^{\infty}\frac{du}{(a^2+u)\Delta}
- \frac{3m}{2} \, c^2
\int\limits_0^{\infty}\frac{du}{(c^2+u)\Delta} \: .
\end{equation}

We linearize equations (\ref{DD-a}) and (\ref{DD-b}) around the
equilibrium configuration of the Maclaurin spheroid ($a = b, \; k
= c/a$). Subtracting (\ref{DD-a}) and (\ref{DD-b}) with using equilibrium values (\ref{ravn-eps}) and (\ref{ravn-Mom}), we obtain
\begin{equation}
\delta(\ddot{a}-\ddot{b}) = 3 \frac{m}{a^3} \,\left[ - k^2
\int\limits_0^{\infty} \frac{du}{(1+u)(k^2+u)^{3/2}} \, +
\int\limits_0^{\infty} \frac{du}{(1+u)^3(k^2+u)^{1/2}} \right]
\delta(a-b).
\end{equation}
The $\varepsilon$, $U_{tot}$, $m$ and $M$ losses are quadratic to
perturbations, so these values remain constant in linear
approximation.

Taking $\delta(a-b) \sim \exp(-i \omega t)$ we come
to the characteristic equation
\begin{equation}
\label{b10}
\omega^2 = 3 \frac{m}{a^3} \,\left[k^2
\int\limits_0^{\infty} \frac{du}{(1+u)(k^2+u)^{3/2}} \, -
\int\limits_0^{\infty} \frac{du}{(1+u)^3(k^2+u)^{1/2}} \,\right].
\end{equation}
These integrals are expressed in analytical functions:
\begin{equation}
\label{b10-1} \omega^2 = \frac{3 m}{a^3} \,\left[k^2 \left(\frac{2}{k(1-k^2)} - \frac{2 \arccos
k}{(1-k^2)^{3/2}} \right) - \left(\frac{3 \arccos k}{4 (1-k^2)^{5/2}} - \frac{3 k}{4 (1-k^2)^2} -
\frac{k}{2(1-k^2)}\right)
 \,\right].
\end{equation}
For the border of stability $\omega^2 = 0$, we obtain the equation again the equation (\ref{b9}).

The spheroid loses its stability for the transformation
into three-axial ellipsoid, at the bifurcation point $k=0.58272$,
$e=\sqrt{1 - c^2/a^2} = 0.81267$. Our approximate equilibrium equations, even without dissipation, describe
the uniformly rotating ellipsoids (spheroids), which are not
connected by adiabatic relations, and contain a "hidden"
non-conservation of the local angular momentum, which preserves
the uniformity. Therefore, in presence of this "hidden"
non-conservation, the loss of stability takes place exactly in the
bifurcation point. The loss of stability is not connected with the
relaxation and follows from our equations even in the case when
all global integrals remain constant. In the exact approach the
instability in this point happens only at the direct presence of
dissipative terms \citep{cha}. The pure adiabatic spheroidal
system preserves stability until $e=0.952887$, where it becomes
unstable via a vibrational mode.

According to the hypothesis of \citet{op73}, the stability of an isolated
axially symmetric system is determined by the ratio $U_{rot}/|U_g|$.
They determined from numerical experiments the critical value for
various configuration as $0.14 \pm 0.03$. It was found in \cite{lrs1} and \citet{shap04} 
that compressible spheroids become secularly unstable to triaxial
deformations at the bifurcation point, where $U_{rot}/|U_g| =
0.1375$, independent of the adiabatic index $n$, $\gamma=1+\frac{1}{n}$. Our formula gives exactly the same
result, which is also confirmed by our numerical simulations.

\section{Emission of very long gravitational waves and gravitational lensing by gravitational waves}

Gravitational radiation is produced during the collapse of the
non-spherical body. Gravitational radiation during a formation of
a pancake was estimated by \cite{Thuan1974}, and was
improved by \cite{Novikov1975}, who took into account the most
important stage of the radiation during a bounce. The formula for
the estimation of the total energy emitted during the collapse and
bounce is
\begin{equation}
\label{eq46w}
U_{GW}\approx 0.01 \left(\frac{r_g}{a_{eq}}\right)^{7/2}
\left(\frac{a_{eq}}{c_{min}}\right)\,Mc^2.
\end{equation}
Here $a_{eq}$ is the equilibrium equatorial radius of the pancake,
and $c_{min}$ is its minimal half-thickness during the bounce,
$r_g=2GM/c^2$ is the Schwarzschild gravitational radius of the body.
 From our calculations we have $a_{eq}/c_{min} \le 100$. The value of
 ${r_g}/{a_{eq}}$ we estimate using the observed average velocity
 of a galaxy in  the rich cluster $v_g \sim 3000$ km/s, and taking
\begin{equation}
\label{eq47w}
\frac{r_g}{a_{eq}} \sim \left(\frac{v_g}{c}\right)^2 \approx 10^{-4}.
\end{equation}
Than the energy carried away by the gravitational wave (GW) may be estimated as
$U_{GW} \approx 10^{-14} M\,c^2$. If all dark matter had passed through the
stage of a pancake formation, than very long GW with a wavelength of the
order of the size of the galactic cluster have an average energy density in the universe
\begin{equation}
\label{eq48w}
\varepsilon_{GW} \approx 10^{-14} \rho_{dm} c^2\approx 3\cdot 10^{-23} {\rm erg/cm}^3.
\end{equation}
Here we have used for estimation the average dark matter density
$\rho_{dm}=3\cdot 10^{-30}$ g/cm$^3$. The average strength
$E_{GW}$ of the very long GW may be estimated, taking the relation
\citep{ll93}
\begin{equation}
\label{eq49w}
\varepsilon_{GW}=\frac{c^2}{16\pi G}{\dot h}^2,
\end{equation}
where $ h $ is metric tensor perturbation (non-dimensional),
connected with GW, we consider only the scalar, having in mind the
averaged value of this perturbation. Taking into account ${\dot
h}=\omega h=2\pi c h/\lambda$,\,  $\lambda\sim 10$ Mpc is the
wavelength of GW of the order of the size of the cluster. From the
comparison of  (\ref{eq48w}),(\ref{eq49w}) we obtain the averaged
metric perturbation due to very long GW in the form \citep{BK2004}
\begin{equation}
\label{eq50w}
\bar h=\frac{2\lambda}{c^2}\left(\frac{G\varepsilon_{GW}}{\pi}\right)^{1/2} \approx 6\cdot 10^{-11}
\end{equation}
for the values of $\lambda$ and $\varepsilon_{GW}$, mentioned
above. Incidently we may expect 10 times larger amplitude of the
GW, than the averaged over the volume value.

According to general relativity, any gravitational field can
change trajectory of photons or, in other words, deflect light
rays. Hence the gravitational field may act as a gravitational
lens.

Gravitational lensing by gravitational waves in different cases
was considered by many authors (see \cite{Nov90}, \cite{Far1992},
\cite{DamEsp} and references therein). It was found that the
deflection angle vanishes for any localized gravitational wave
packet because of transversality of gravitational waves
\citep{DamEsp}. Thus if the photon passes through the finite
gravitational wave pulse its deflection due to this wave is equal
to zero. In \cite{BKTs2008b} we notice that the displacement between
trajectories of the photon before and after passing the wave may
occur. On the basis of this
result we obtain an approximate formula for the estimation of
observational effects.

\begin{figure}
\centerline{\hbox{\includegraphics[width=0.4\textwidth]{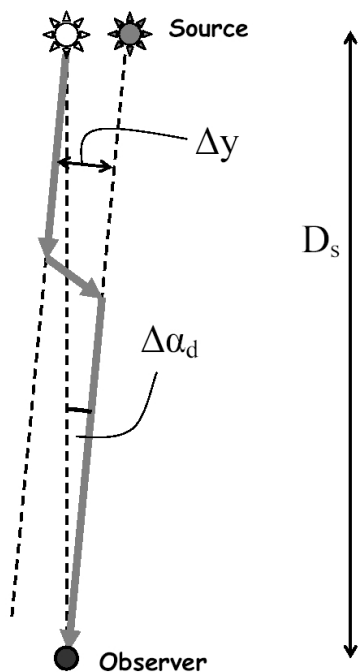}}}
\caption{The observational effect of the displacement in
trajectory of the photon.}
\label{fig-wave3}
\end{figure}

We confirm analytically vanishing of deflection angle
for plane wave pulses. However, we have found that the
gravitational wave (GW) changes the photon propagation in another
way, simply shifting its whole trajectory after passing through
the GW (see fig. \ref{fig-wave3}). This displacement is found analytically for
the photon passing through the plane GW. Displacement takes place mainly in the case of isolated wave
pulses, which have a form similar to the top part of the sinusoid
or when it has the non-symmetrical top and bottom parts of wave
profile, and may, in principal, vanish for the periodic wave of a
long duration.

Directions of photons passing through the gravitational wave
packet does not change, therefore any focusing of rays does not
occur in this case. Thus the displacement in trajectories does not
lead to any magnification effect. But the displacement leads to
change of the angular position of object for distant observer. The
change of the angular position due to passing of the light ray
through the gravitational wave pulse $\Delta \alpha_d$ is as (see
fig. \ref{fig-wave3})
\begin{equation}
\Delta \alpha_d = \frac{\Delta y}{D_s} \simeq \frac{h \delta}{D_s}
\, ,
\end{equation}
where $h$ is the amplitude of the GW pulse, $\delta$ is its
thickness and $D_s$ is a distance between the source and the
observer.

Let us estimate the change of the angular position for the GW
pulses produces during formation of large scale structure of the
Universe in dark matter using the following parameters:
$h = 10^{-11}$, $\delta = Mpc$, $D_s = 100 Mpc$. Than we obtain
\begin{equation}
\Delta \alpha_d \simeq 2 \cdot 10^{-8} \, \mathrm{arcsec}.
\end{equation}

Another possible manifestation of gravitational waves is influence of them on the polarization of cosmic microwave background \citep{Polnarev1985}. It is known that primordial GWs lead to formation of B-mode polarization \citep{SeljakZ1997}, see also textbook of \cite{Gorbunov2011}. This effect is now hunting by different observational projects, e.g. see BICEP2 (http://bicepkeck.org/). We suppose that B-mode polarization formation can be explained not only by primordial gravitational waves but also by gravitational waves appeared in the epoch of the formation of the large scale structure in the universe. The main distinguishing feature of gravitational waves formed during large scale structure formation is anisotropy in direction to rich clusters of galaxies.

\section{Conclusions}

We present here the approximate way for investigation of collapse and dynamic stability based on the model of compressible homogeneous 3-axis ellipsoids. Investigation of dynamics is reduced to numerical solution of system of ordinary differential equations of semi-axes time evolution. In the presence of dissipation processes there are also equations for mass, angular momentum, energy, and entropy evolution. Derivation of equations of motion have been performed by variation of the Lagrange function of the ellipsoid. Different equations of state can be easily considered in this model. Right-hand sides of equations have analytical form for spheroid case and can be written in terms of elliptical integrals for 3-axis ellipsoid case. Equilibrium configurations can be also investigated by taking zero accelerations in all equations. We apply this method for investigation of dynamics of non-spherical stars and dark matter large scale structure formation.

Dynamic stability of non-spherical stars was investigated. We have solved the equations that describe, in a simplified way, the Newtonian dynamics of a self-gravitating non-rotating spheroidal body after loss of stability. We have found that contraction to a singularity occurs only in a pure spherical collapse, and deviations from spherical symmetry
stop the contraction through the stabilizing action of non-linear non-spherical oscillations. A
real collapse occurs after damping of the oscillations because of energy losses, shock wave
formation or viscosity.

We have investigated the dynamics of 3-axis dark matter ellipsoid.
Equations of motion for axes of uniform compressible ellipsoid
have been obtained by variation of the Lagrange function, in which
the violent relaxation and losses of a matter, energy and angular
momentum have been included phenomenologically.

The system was solved numerically, until the formation of
stationary rotating figures in presence of the relaxation. For
lower angular momentum $M$ we have a formation of a compressed
spheroid, while at larger $M$ we follow the development of
three-axial instability and formation of three-axial ellipsoid.
The instability in this approximation happens at the bifurcation
point of the sequence of Maclaurin spheroids, where Jacobi
ellipsoidal system starts.

The bifurcation point coinciding with the point of loss of
stability is found analytically in the form of a simple formula,
by static and dynamic approaches. Numerical and analytical
considerations give identical results. Development of instability, connected with radial orbits, is obtained for slowly rotating collapsing bodies.

The weak but very long gravitational waves (GW) emitted mainly on
the stages of the collapse and pancake formation form a long wave
GW background. The existence of such a long GW in the space between the source and the observer may be registered as an action of the gravitational lens. Effect of lensing can be due to displacement of the whole trajectory if light ray is passing through the GW pulse. The displacement leads to
change of the angular position of object for distant observer.

The challenging idea is explanation of B-mode polarization formation not only by primordial gravitational waves but also by gravitational waves appeared in the epoch of the formation of the large scale structure in the universe.

\section*{Acknowledgements}

The work of GSBK and OYuT was partly supported by the Russian Science Support Foundation, Russian Basic Research Foundation Grant No. 14-02-00728 and the Russian Federation President Grant for Support of Leading Scientific Schools, Grant No. NSh-261.2014.2.

The work of GSBK was also partially supported by the Russian Foundation for Basic Research Grant No. OFI-M 14-29-06045.


\bibliographystyle{jpp}


\begin{thebibliography}{99}


\bibitem[\protect\citeauthoryear{Antonov}{1973}]{ant} Antonov V.A., 1973,
The Dynamics of Galaxies and Stellar Clusters. Nauka, Alma-Ata (in
Russian)

\bibitem[\protect\citeauthoryear{Ardeljan, Bisnovatyi-Kogan
\& Moiseenko}{2005}]{abkm05}
Ardeljan N.V., Bisnovatyi-Kogan G.S., Moiseenko S.G., 2005, Magnetorotational supernovae, \textit{MNRAS}, \textbf{359}, 333--344

\bibitem[\protect\citeauthoryear{Bisnovatyi-Kogan}{1989}]{BK1989}
Bisnovatyi-Kogan G.S., 1989, Physical problems in the theory of
stellar evolution. Nauka, Moscow (in Russian). (English
translation: Stellar Physics, Vol. 1,2. Springer, 2001)

\bibitem[\protect\citeauthoryear{Bisnovatyi-Kogan}{2004}]{BK2004}
Bisnovatyi-Kogan G.S., 2004, A simplified model of the formation of structures in dark matter and a background of very long gravitational waves, \textit{MNRAS} \textbf{347}, 163--172

\bibitem[\protect\citeauthoryear{Bisnovatyi-Kogan \& Tsupko}{2005}]{BKTs2005}
Bisnovatyi-Kogan G.S. and Tsupko O.Yu., 2005, Approximate dynamics of dark matter ellipsoids, \textit{MNRAS} \textbf{364}, 833--842

\bibitem[\protect\citeauthoryear{Bisnovatyi-Kogan \& Tsupko}{2008a}]{BKTs2008}
Bisnovatyi-Kogan G.S. and Tsupko O.Yu., 2008a, Dynamic stabilization of non-spherical bodies against unlimited collapse, \textit{MNRAS} \textbf{386}, 1398--1403


\bibitem[\protect\citeauthoryear{Bisnovatyi-Kogan and Tsupko}{2008b}]{BKTs2008b}
Bisnovatyi-Kogan G.S. and Tsupko O.Yu., 2008b, Gravitational lensing by gravitational waves, \textit{Gravitation and Cosmology},  V.\textbf{14}, N.3, 226--229.


\bibitem[\protect\citeauthoryear{Boily et al}{1999}]{Boily1999}
Boily C.M., Clarke C.J., Murray S.D. 1999, Collapse and evolution of flattened star clusters, \textit{Monthly Notice of the Royal Astronomical Society}, \textbf{302}, 399.

\bibitem[\protect\citeauthoryear{Boily et al}{2002}]{Boily2002}
Boily C.M., Athanassoula E., Kroupa P., 2002, Scaling up tides in numerical models of galaxy and halo formation, \textit{Monthly Notice of the Royal Astronomical Society} \textbf{332}, 971.

\bibitem[\protect\citeauthoryear{Braginsky et al}{1990}]{Nov90}
Braginsky V.B., Kardashev N.S., Polnarev A.G., and Novikov I.D., 1990, Propagation
of electromagnetic radiation in a random field of gravitational waves and space radio
interferometry. \textit{Nuovo Cimento} \textbf{B 105}, 1141.

\bibitem[\protect\citeauthoryear{Chandrasekhar}{1969}]{cha}
Chandrasekhar S., 1969, Ellipsoidal Figures of Equilibrium. Yale
Univ. Press, New Haven


\bibitem[\protect\citeauthoryear{Damour and Esposito-Far\`{e}se}{1998}]{DamEsp}
Damour T., Esposito-Far\`{e}se G., 1998, Light deflection by gravitational waves from localized sources, \textit{Phys. Rev.} \textbf{D}
\textbf{58}, 042003.

\bibitem[\protect\citeauthoryear{Doroshkecich et al}{1980}]{Doroshkevich1980}
Doroshkevich A.G., Kotok E.V., Polyudov A.N., Shandarin S.F., Sigov Yu.S., Novikov I.D. 1980, Two-dimensional simulation of the gravitational system dynamics and formation of the large-scale structure of the universe, \textit{MNRAS}, \textbf{192}, 321--327

\bibitem[\protect\citeauthoryear{Doroshkevich et al}{1999}]{Doroshkevich1999}
Doroshkevich A.G., M\"uller V., Retzlaff J., Turchaninov V. 1999, Superlarge-scale structure in N-body simulations, \textit{MNRAS} \textbf{306}, 575--591

\bibitem[\protect\citeauthoryear{Faraoni}{1992}]{Far1992}
Faraoni V., 1992, Nonstationary gravitational lenses and the Fermat principle, \textit{Astrophys. J.} \textbf{398}, 425--428.

\bibitem[\protect\citeauthoryear{Fridman \& Polyachenko}{1985}]{fp85}
Fridman A.M., Polyachenko V.L., 1985, Physics of Gravitating
Systems. Springer Verlag, Berlin

\bibitem[\protect\citeauthoryear{Gorbunov \& Rubakov}{2011}]{Gorbunov2011}
Gorbunov D.S., Rubakov V.A. 2011 Introduction to the Theory of the Early Universe. Cosmological Perturbations and Inflationary Theory. World Scientific Publishing, Singapore

\bibitem[\protect\citeauthoryear{Klypin \& Shandarin}{1983}]{Shandarin1983}
Klypin A. A., Shandarin S. F., 1983, Three-dimensional numerical model of the formation of large-scale structure in the Universe, \textit{Monthly Notices of the Royal Astronomical Society}, \textbf{204}, 891-907


\bibitem[\protect\citeauthoryear{Lai, Rasio \& Shapiro}{1993}]{lrs1}
Lai D., Rasio F.A., Shapiro S.L., 1993, Ellipsoidal figures of equilibrium - Compressible models, \textit{ApJS}, \textbf{88}, 205

\bibitem[\protect\citeauthoryear{Lai, Rasio \& Shapiro}{1994a}]{lrs4}
Lai D., Rasio F.A., Shapiro S.L., 1994a, Equilibrium, stability, and orbital evolution of close binary systems \textit{ApJ}, \textbf{423}, 344.

\bibitem[\protect\citeauthoryear{Lai, Rasio \& Shapiro}{1994b}]{lrs5}
Lai D., Rasio F.A., Shapiro S.L., 1994b, Hydrodynamics of rotating stars and close binary interactions: compressible ellipsoid models, \textit{ApJ}, \textbf{437}, 742.

\bibitem[\protect\citeauthoryear{Landau \& Lifshitz}{1993}]{ll93}
Landau L.D., Lifshitz E.M., 1993, The Classical Theory of Fields.
Pergamon, Oxford

\bibitem[\protect\citeauthoryear{Lichtenberg \& Lieberman}{1983}]{LL83}
Lichtenberg A.J., Lieberman M.A., 1983, Regular and Stochastic motion.
Springer-Verlag, New York

\bibitem[\protect\citeauthoryear{Lin, Mestel \& Shu}{1965}]{LMS}
Lin C.C., Mestel L., Shu F.H., 1965, The Gravitational Collapse of a Uniform Spheroid, \textit{ApJ}, \textbf{142}, 1431.

\bibitem[\protect\citeauthoryear{Lynden-Bell}{1964}]{LB64}
Lynden-Bell D., 1964, On large-scale instabilities during gravitational collapse and the evolution of shrinking Maclaurin spheroids, \textit{ApJ}, \textbf{139}, 1195.

\bibitem[\protect\citeauthoryear{Lynden-Bell}{1965}]{LB65}
Lynden-Bell D., 1965, On the evolution of frictionless ellipsoids, \textit{ApJ}, \textbf{142}, 1648.

\bibitem[\protect\citeauthoryear{Lynden-Bell}{1967}]{ref2}
Lynden-Bell D., 1967, Statistical mechanics of violent relaxation in stellar system, \textit{MNRAS}, \textbf{136}, 101.


\bibitem[\protect\citeauthoryear{Lynden-Bell}{1996}]{LB96}
Lynden-Bell D., 1996, Consequences of one spring researching with Chandrasekhar \textit{Current Science}, \textbf{70}, 789.

\bibitem[\protect\citeauthoryear{Mazets et al.}{1981}]{mg81}
Mazets E.P., Golenetskii S.V., Il'inskii V.N., Panov V. N., Aptekar R.L., Gur'yan Y.A., Proskura M.P., Sokolov I.A., Sokolova Z.Ya., Kharitonova T.V., 1981, Catalog of cosmic gamma-ray bursts from the KONUS experiment data, \textit{Ap\&SS}, \textbf{80}, 3, 85, 109

\bibitem[\protect\citeauthoryear{ Moiseenko, Bisnovatyi-Kogan
\& Ardeljan}{2005}]{mbka06}
Moiseenko S.G., Bisnovatyi-Kogan G.S., Ardeljan N.V., 2006, A magnetorotational core-collapse model with jets, \textit{MNRAS}, \textbf{370}, 501--512

\bibitem[\protect\citeauthoryear{Novikov}{1975}]{Novikov1975}
Novikov I.D., 1975, Gravitational radiation from a star that is contracting into a disk. \textit{Astron. Zh.}
\textbf{52}, 657?659 (in Russian); English translation: Novikov, I. D., 1976, Gravitational radiation
from a star collapsing into a disk. \textit{Sov. Astron.} \textbf{19}, 398?399..

\bibitem[\protect\citeauthoryear{Ostriker \& Peebles}{1973}]{op73}
Ostriker J.P., Peebles P.J.E., 1973, A numerical study of the stability of flattened galaxies: or, can cold galaxies survive? \textit{ApJ}, \textbf{186}, 467

\bibitem[\protect\citeauthoryear{Polnarev}{1985}]{Polnarev1985}
Polnarev A. G., 1985, Polarization and anisotropy induced in the microwave background by cosmological gravitational waves, \textit{Sov. Astron.} \textbf{29}, 607.


\bibitem[\protect\citeauthoryear{Rodrigues}{2014}]{Rodrigues2014}
Rodrigues H., 2014, On determining the kinetic content of ellipsoidal configurations,
\textit{MNRAS} \textbf{440}, 1519--1526

\bibitem[\protect\citeauthoryear{Rosensteel \& Tran}{1991}]{RT91}
Rosensteel G., Tran H.Q., 1991, Hamiltonian dynamics of self-gravitating ellipsoids, \textit{ApJ}, \textbf{366}, 30




\bibitem[\protect\citeauthoryear{Seljak \& Zaldarriaga}{1997}]{SeljakZ1997}
Seljak U., Zaldarriaga M., 1997, Signature of Gravity Waves in the Polarization of the Microwave Background, \textit{Phys. Rev. Lett.} \textbf{78}, 2054.


\bibitem[\protect\citeauthoryear{Shandarin \& Zeldovich}{1989}]{Shandarin1989}	
Shandarin S. F., Zeldovich, Ya. B. 1989 The large-scale structure of the universe: Turbulence, intermittency, structures in a self-gravitating medium, \textit{Reviews of Modern Physics}, \textbf{61}, 185-220.



\bibitem[\protect\citeauthoryear{Shapiro}{2004}]{shap04}
Shapiro S.L., 2004, The secular bar-mode instability in rapidly rotating stars revisited, \textit{ApJ}, \textbf{613}, 1213.

\bibitem[\protect\citeauthoryear{Sharif \& Zaeem Ul Haq Bhatti}{2013}]{Sharif2013}
Sharif M. and Zaeem Ul Haq Bhatti M., 2013, Stability analysis of restricted non-static axial symmetry, \textit{Journal of Cosmology and Astroparticle Physics}, Issue \textbf{11}, article id. 014

\bibitem[\protect\citeauthoryear{Sharif \& Zaeem Ul Haq Bhatti}{2014}]{Sharif2014}
Sharif M. and Zaeem Ul Haq Bhatti M., 2014, On the stability of a class of radiating viscous self-gravitating stars with axial symmetry, \textit{Astroparticle Physics}, \textbf{56}, 35--41

\bibitem[\protect\citeauthoryear{Sneth at al}{2001}]{Sneth2001}
Sheth R.K., Mo H.J., Tormen G. 2001 Ellipsoidal collapse and an improved model for the number and spatial
distribution of dark matter haloes, \textit{Monthly Notice of the Royal Astronomical Society}, \textbf{323}, 1.

\bibitem[\protect\citeauthoryear{Thuan and Ostriker}{1974}]{Thuan1974}
Thuan T.X., Ostriker J.P. 1974, Gravitational Radiation from Stellar Collapse, \textit{ApJ}, \textbf{191}, L105--L107.

\bibitem[\protect\citeauthoryear{Vandervoort}{2011}]{vand2011}
Vandervoort P.O., 2011, On chaos in the oscillations of galaxies,
\textit{Mon. Not. R. Astron. Soc.} \textbf{411}, 37--53

\bibitem[\protect\citeauthoryear{Vandervoort}{2014}]{vand2014}
Vandervoort P.O., 2014, On chaos in the pulsations of stars, \textit{MNRAS} \textbf{443}, 504--521


\bibitem[\protect\citeauthoryear{Zeldovich}{1964}]{z1964}
Zeldovich Ya. B., 1964, Newtonian and Einsteinian Motion of Homogeneous Matter, \textit{Astronomicheskii Zhurnal}, \textbf{41}, 873; English translation: Zeldovich Ya. B., 1965, Newtonian and Einsteinian Motion of Homogeneous Matter, \textit{Soviet Astronomy}, \textbf{8}, 700

\bibitem[\protect\citeauthoryear{Zeldovich}{1970a}]{z70a}
Zeldovich Ya.B., 1970a, Separation of uniform matter into parts under the action of gravitation, \textit{Astrofizika}, \textbf{6}, 319-335; English translation: Zeldovich Ya.B., 1970, Fragmentation of a homogeneous medium under the action of gravitation, \textit{Astrophysics}, \textbf{6}, 164--174


\bibitem[\protect\citeauthoryear{Zeldovich}{1970b}]{z70b}
Zeldovich Ya.B., 1970b, Gravitational instability: An approximate theory for large density perturbations, \textit{Astronomy and Astrophysics},	
\textbf{5}, 84-89


\bibitem[\protect\citeauthoryear{Zeldovich \& Novikov}{1967}]{ZN}
Zeldovich Ya.B., Novikov I.D., 1967, Relativistic Astrophysics.
Nauka, Moscow (in Russian)














\end{thebibliography}

\end{document}